\documentclass[12pt]{article}
\usepackage[top=2.5cm,bottom=2.5cm,left=2cm,right=2cm]{geometry}  
\usepackage{graphicx}   
\usepackage{xcolor}      

\usepackage[english]{babel} 
\usepackage[T1]{fontenc}    
\usepackage[ansinew]{inputenc} 
\usepackage{lmodern} 
\usepackage[condensed,light,math]{iwona}
\usepackage{mathtools}  
\usepackage{scalerel}
\def\lVert{\mid\!\mid}
\def\rVert{\mid\!\mid}
\usepackage{microtype} 

\usepackage{amsmath} 
\usepackage{amssymb} 
\usepackage{algorithm}  
\usepackage{algpseudocode}

\usepackage[switch,running,right,mathlines]{lineno}  
\usepackage{csquotes} 
\usepackage[style=authoryear,backend=biber,giveninits=true,maxcitenames=2,maxbibnames=99,urldate=edtf,seconds=true,uniquename=false,uniquelist=false,url=true]{biblatex}

\usepackage[matstyle=bbold]{MattsMacros}
\usepackage{hyperref}  

\graphicspath{{./images/}}

\title{Bayesian Functional Generalized Additive Models with Sparsely Observed Covariates}

\author{Mathew W.\ McLean\thanks{School of Mathematical and Physical
    Sciences, University of Technology Sydney, P.O. Box 123, Ultimo,
    NSW 2007, Australia (E-mail: mathew.w.mclean@gmail.com)}
  \and
  Fabian Scheipl \thanks{Research Associate, Department of Statistics,
    Ludwig-Maximilians-University of Munich, 80539, Munich, Germany,
    (Email: fabian.scheipl@stat.uni-muenchen.de)}
  \and
  Giles Hooker\thanks{Associate Professor, Department of Biological
    Statistics and Computational Biology, Cornell University, Ithaca,
    NY, 14853, USA (E-mail: giles.hooker@cornell.edu)}
  \and
  Sonja Greven \thanks{Emmy Noether Junior Research Group Leader, Department
    of Statistics, Ludwig-Maximilians-University of Munich, 80539
    Munich, Germany (Email: sonja.greven@stat.uni-muenchen.de)}
  \and
  David Ruppert \thanks{Andrew Schultz Jr.\ Professor of Engineering
    and Professor of Statistical Science, School of Operations
    Research and Information Engineering and Department of Statistical
    Science, Cornell University, 1170 Comstock Hall, Ithaca, NY 14853,
    USA (E-mail: dr24@cornell.edu) } }

\date{\today}

\begin{document}
\renewcommand{\linelabel}[1]{}
\date{\today}
\maketitle
\begin{abstract}  
  \linelabel{editor:4b}We propose semiparametric Bayesian methods for
  scalar-on-function regression involving sparse longitudinal data.
  Our work extends the functional generalized additive model (FGAM) of
  \textcite{McLean2012functional}, a recently proposed model offering
  greater flexibility than the common functional linear model (FLM).
  The algorithms we develop allow for the functional covariates to be
  sparsely observed and measured with error, whereas the estimation
  procedure of \textcite{McLean2012functional} required that they be
  noiselessly observed on a regular grid.  The Bayesian approaches we
  present estimate the functional predictors simultaneously with all
  other model parameters, and hence automatically account for
  variability in the estimated predictors, which is not possible with
  current frequentist approaches in the literature.  We consider both
  Monte Carlo and variational Bayes methods for fitting the FGAM with
  sparsely observed covariates.  Due to the complicated form of the
  model posterior distribution and full conditional distributions,
  standard Monte Carlo and variational Bayes algorithms cannot be
  used.  The strategies we use to handle the updating of parameters
  without closed-form full conditionals should be of independent
  interest to applied Bayesian statisticians working with nonconjugate
  models.  Our numerical studies demonstrate the benefits of our
  algorithms over a two-step approach of first recovering the complete
  trajectories using standard techniques and then fitting a functional
  regression model.  Our methods are applied to forecasting closing
  price for items up for auction on the online auction website eBay.
\end{abstract}
\baselineskip=18pt
\section{Introduction}
\linelabel{editor:4c}In this work, we extend a general class of models
for functional regression to the longitudinal data setting, where each
sampled function may have only a small number of noisy measurements at
irregular time intervals.  This setting presents considerable
challenges which cannot be handled by conventional estimation methods
in the functional regression literature.  Previous work in this area
assumes a linear relationship between the response and covariate,
which is often inappropriate.  We build on the work of
\textcite{McLean2012functional}, in order to allow for more general
response-predictor relationships.  The estimation methods used in
\textcite{McLean2012functional} depend upon the functional data being
fully observed without error.  A naive approach of simply
interpolating between points for each functional covariate, and then
assuming the trajectories are completely observed can be very
inaccurate, and we therefore must develop an entirely different
approach to \textcite{McLean2012functional}.  We will take a Bayesian
approach, which will allow us to simultaneously recover the complete
functional trajectories while estimating all other model parameters.

It is now commonplace in many fields to collect data where each
observation is a sample path from some underlying continuous-time
stochastic process, $\{X(t): t\in \mcT\}$.  Functional data analysis
(FDA) is the branch of statistics concerned with methods for analyzing
such data.  FDA methods often rely on an assumption of smoothness of
the underlying process and ordinarily assume the sampled trajectories
$X_i(t)$ are fully and noiselessly observed.  Typically, the $X(t)$
are represented as the result of some presmoothing of the data.

One problem that is frequently studied in the FDA literature is that
of using the sampled trajectories as covariates in a regression model
involving a scalar response variable.  The most commonly used model in
this setting is the functional linear model (FLM), first proposed in
\textcite{ramsay1991some}, given by \beq E(Y_i\mid
X_i)=\theta_0+\int_\mathcal{T}\beta(t)X_i(t)\,dt,\quad
i=1,\ldots,N;\label{flm} \eeq where $X_i$ is a real-valued,
continuous, square-integrable, random curve on the compact interval
$\mcT$, $Y_i$ is a scalar random variable, $\theta_0$ an intercept,
and $\beta(\cdot)$ is the functional coefficient with $\beta(t)$
describing the effect on the response of the functional predictor at
time $t$.

A model recently proposed in \textcite{McLean2012functional} called
the functional generalized additive model (FGAM) removes the
restrictive linearity assumption of the FLM by modeling the
conditional mean of $Y$ as \beq E(Y_i\mid X_i) = \eta_{0i} + \int_\mcT
F\{X_i(t),t\}\,dt,\label{fgam} \eeq where $F$ is an unknown smooth
function and the offset term $\eta_{0i}$ contains any additional
scalar or functional covariates other than $X_i(t)$.  Notice that as a
special case, when $F\{X(t),t\}=\beta(t)X(t)$ and
$\eta_{0i}=\theta_0$, we obtain the FLM.  This model retains the ease
of interpretability of the FLM while suffering from less approximation
bias.  The surface $F$ will be parameterized using tensor products of
B-splines and two smoothing parameters will control the complexity of
the estimated surface.  As the FLM can be thought of as a
(multivariate) linear model with an infinite number of predictors, the
FGAM can be thought of as an additive model in an infinite number of
predictors \autocite{McLean2012functional}.

Frequently, the functional data we encounter in practice are not
observed on a dense, regularly-spaced grid, but instead on a sparse,
irregular grid with measurement error and with some subjects having as
little as one or two measurements.  This type of data is frequently
found in the longitudinal data analysis (LDA) literature.  An overview
of the differences between FDA and LDA can be found in
\textcite{rice2004functional}.  When the trajectories are not observed
on a regular grid, the estimation procedure used in
\textcite{McLean2012functional} cannot be directly applied.  In these
situations, the semiparametric techniques commonly used in LDA are
more appropriate; in this work, we take a linear mixed effects
modeling approach.  Most of the previous work on sparsely observed
functional data only considers estimation of the mean and covariance
function of the underlying process, with few papers examining
regression of a scalar on the sparse trajectories.  Notable exceptions
are \textcite{james2002generalized}, \textcite{wang2005efficient},
\textcite{bigelow2009bayesian}, and \textcite{goldsmith2011penalized}.
To the best of our knowledge, our work is the first to study
nonparametric regression with sparse functional data.

It is common to estimate the complete functional trajectories by
performing a functional principal components analysis (FPCA); for
example, the principal components analysis through conditional
expectation (PACE) method of \textcite{yao2005functional}.  Whereas a
typical functional data analysis smooths the measurements for each
subject separately, the advantage of PACE is that it pools data across
subjects at each time point to estimate an entire covariance surface.
This ``borrowing of strength'' across subjects is a main reason for
the method's success.  Although it is not considered in
\textcite{yao2005functional}, one might think it reasonable to use a
two-stage approach of first using PACE to recover the function
predictors and then in a second step fitting an FLM using standard
techniques or an FGAM using the procedure in
\textcite{McLean2012functional}.  The main advantage of our Bayesian
algorithms over a two-stage approach is that they allow us to directly
account for uncertainty in the estimates from the FPCA. Our numerical
results demonstrate the inadequacy of a conventional two-stage
estimation procedure and we believe that our algorithms also gain from
using information in the response when estimating the functional
trajectories.

An important step in the PACE procedure is estimating the covariance
surface of the functions using local polynomial modeling.  Although
PACE often performs well in a variety of situations, in our simulation
studies we observe similar results to \textcite{peng2009geometric},
who found that PACE can have problems in more challenging settings
with higher sparsity and a true covariance function that has more than
three non-zero eigenvalues.  In a number of the simulations in
\textcite{peng2009geometric}, and in our own experiments, the
covariance surface estimated by PACE is not positive definite and the
estimated measurement error variance is negative.  We will demonstrate
that our Bayesian algorithms do not suffer from this problem.  Our
methods can also be used to effectively recover a greater number of
principal components.  Several currently available techniques only
consider recovery of two non-zero principal components in simulation
studies and attempt to estimate three components in real data studies
\autocite[e.g.,][]{yao2005functional,yao2005penalized}.

Our goals are three-fold: 1) accurate recovery of the sparsely
observed trajectories, 2) accurate recovery of the surface, $F(x,t)$,
and 3) accurate prediction of the response, $Y$.  The missing parts of
the trajectories must be imputed during the estimation procedure.
Three possibilities for doing this are an expectation-maximization
(EM) algorithm, Markov Chain Monte Carlo (MCMC), or a variational
approximation.  The advantage of MCMC over an EM algorithm approach is
that uncertainty about the imputed curves is automatically taken into
account during the estimation.  Due to the computational overhead
associated with MCMC, we also present a variational Bayes algorithm
that can be used for fast approximate inference and to initialize an
MCMC sampler.

Variational Bayes (VB) refers to a specific variational approximation
used for Bayesian inference that relies on the assumption that a
posterior density of interest factors into a product form over certain
groups of model parameters.  Though they are commonly used in computer
science, the application of variational approximations in statistics
is relatively new; \textcite{ormerod2010explaining} provides an
overview.  When the amount of posterior dependence is small, there is
little loss of accuracy and often very large improvements in
computation time over MCMC methods.  Applications of VB to regression
problems with missing data can be found in
\textcite{faes2011variational} and \textcite{goldsmith2011functional},
the latter of which considered the FLM.

The success of the approximation hinges on the amount of between-group
dependence among the parameters in the posterior distribution.  The
cost of the computational efficiency gains from the approximations
made in VB is the loss of guaranteed convergence to the correct
distribution provided by MCMC.  Factorization assumptions are often
reasonable for certain groups of parameters in functional data models
\autocite{goldsmith2011functional}.  We agree with those authors that
VB should not be considered a replacement for fully Bayesian
inference.  Instead we consider it as complementary to MCMC: a useful
tool for approximate answers in large data situations when MCMC
becomes intractable.  One natural way to use the two as complements is
to use VB estimates as starting values for an MCMC algorithm in the
hopes of achieving faster convergence to, and better exploration of,
the posterior distribution of interest.  In our experience, the choice
of starting values is critical for high-dimensional problems such as
functional regression.

When conjugate priors are used and closed-form expressions exist for
all full conditional distributions in a model, the optimal densities
for approximating the posterior using VB have closed-form expressions
as well.  It is not possible to obtain closed-form updates for all the
paramaters in the FGAM due to the nonconjugate full conditional
distribution for the principal component scores, as they appear in the
likelihood as arguments to the B-spline basis functions used to
parameterize the regression surface.  Therefore, Metropolis-Hasting
steps are needed for our MCMC algorithm.  For our VB algorithm, we
alternatively overcome the nonconjugacy using a Laplace approximation.
An additional complication is the necessity of an anisotropic
roughness penalty for $F(x,t)$, owing to the possibly differing
amounts of smoothness in $x$ and $t$, which makes the two smoothing
parameters difficult to separate.  Using our VB approach, we are
typically able to obtain a speed-up of at least an order of magnitude
over generating 10,000 samples from our MCMC sampler, with minimal
sacrifice in accuracy.  Our approaches perform quite well at both
out-of-sample prediction and recovering the true surface whether the
true model is linear or
nonlinear.

The remainder of the paper proceeds as follows: Section 2 briefly
reviews functional principal component analysis, Section 3 discusses
our parameterization for the unknown surface, $F(x,t)$, Section 4
discusses our MCMC algorithm for fitting FGAM, Section 5 reviews
variational Bayes and provides a VB algorithm for fitting FGAM,
Section 6 discusses results of simulation experiments, in Section 7 we
apply our algorithms to forecasting closing prices for seven day
auctions on the auction website eBay, and Section 8
concludes.
%
\section{Recovering Sparsely Observed Functional Data}\label{PACE}
In this section we give a brief overview of the literature on
estimating trajectories from sparsely observed functional data; one of
our goals mentioned in the previous section and a key step in building
our regression model.  Most methods involve various techniques for
estimating eigenfunctions and eigenvalues from an FPCA.  A common
approach for this is to use mixed model representations for penalized
or smoothing splines; see \textcite{james2000principal} and the
references therein.  Another frequently used approach uses local
polynomial modeling; see e.g., \textcite{yao2005functional}.  Bayesian
approaches to functional data analysis include the wavelet-based mixed
model method of \textcite{morris2006wavelet} and the Dirichlet process
based approach of \textcite{rodriguez2009bayesian}.  Though some
papers in the Bayesian literature, including the ones cited above,
appear to be able to deal with irregularly sampled functional data, it
is unclear how these methods perform in the high-sparsity situations
we wish to consider here, and we are not aware of any of these papers
analyzing how their methods perform under varying degrees of
sparsity/missingness.

The usual model for the unknown functions is to assume $n_i$ noisy
measurements have been taken of $X_i(t)$:
$\tvecx_i=\{\widetilde{x}_i(t_{i,1}),\ldots,\widetilde{x}_i(t_{i,n_i})\}^T$
with
 $\widetilde x_i(t_{ij}) = X_i(t_{ij}) + e_{ij};\ e_{ij} \stackrel{\text{i.i.d.}}{\sim} N(0, \sigma^2_x);\ i=1,\ldots,N;\ j=1,\ldots,n_i.$
 We define the mean and covariance functions $\mu_x(t):=E\{X(t)\}$ and
 $G(s,t):=\text{Cov}\{X(s), X(t)\}$.  If $X\in\mathcal{L}^2$, then by
 Mercer's theorem $G(s,t)$ admits an expansion
 $G(s,t)= \sum^\infty_{m=1} \nu_m \phi_m(s) \phi_m(t)$ with
 (orthonormal) eigenfunctions $\phi_m(\cdot)$ and associated
 eigenvalues $\nu_m$, and the curves have a Karhunen-Lo\`{e}ve
 representation
$X_i(t)=\mu_x(t) + \sum^\infty_{m=1} \phi_m(t) \xi_{im};\ \xi_{im} \simind (0, \nu_{m})$,
where the $\xi$'s are known as principal component (PC) scores.  If
$X(t)$ is assumed to be a Gaussian process, then the principal
component scores are Gaussian random variables.

For all FPCA methods, it is necessary to choose an integer, $M$, at
which to truncate the basis expansion for the unknown functions (i.e.\
assume $\xi_k=0$ for all $k>M$).  This is typically done by including
enough scores to explain a prespecified percentage (e.g. $99\%$) of
the total observed variation in the data, and that is the approach we
take in our analysis of the auction data in
Section~\ref{auction}.

\linelabel{ref2:1}To initialize both our MCMC and VB algorithms, we
take a similar (though not identical) approach to
\textcite{yao2005functional}.  We use P-splines
\autocite{eilers1996flexible} for the smoothing steps 1.\ and 2.\
described below whereas \textcite{yao2005functional} took a local
polynomial modelling approach for these steps.  The use of penalized
splines to perform FPCA is covered in detail by
\textcite{yao2005penalized}.  We perform the smoothing using the
\texttt{R} package \texttt{mgcv} \autocite{wood2006generalized} and
use generalized cross validation to choose smoothing parameters.  The
full list of steps for performing the FPCA are as follows.
\begin{enumerate}
\item Obtain an estimate $\widehat{\mu}(t)$ of $\mu(t)$ via
  semiparametric regression of the pooled data
  $\tvecx = (\tvecx_1\tr,\dots,\tvecx_N\tr)\tr$ on
  $\tilde{\vect} = (\vect_{1}^T,\ldots,\vect_{N}^T)\tr$;
  $\vect_i=(t_{i,1},\ldots,t_{i,n_i})^T$ using P-splines.
\item Obtain an estimate $\hat{G}(s,t)$ of $G(s,t)$ by fitting a cubic
  tensor-product P-spline \autocite{marx2005multidimensional} to the
  ``raw'' covariances with the diagonal removed:
  $\{\widetilde{x}_i(t_{il}) -
  \widehat\mu_x(t_{il})\}\{\widetilde{x}_i(t_{is}) -
  \widehat{\mu}_x(t_{is})\},\ l\neq
  s$.  
  We use third-derivative penalties when fitting the tensor product
  spline in order to shrink estiamte towards a quadratic surface.  We
  write $\hat{G}(\vect_i,\vect_i)$ to denote the $N_i\times
  N_i$ matrix with $(j,k)$-entry $\hat{G}(t_{ij},t_{ik})$.
\item $\sigma_x^2$ is estimated as the average of the middle two thirds of
  the diagonal of the raw covariance matrix minus the diagonal of the
  smoothed covariance surface.  This is as in
  \textcite{yao2005functional} and is done to avoid boundary effects.
\item $\hbnu=(\hat\nu_1,\ldots,\hat\nu_M)^T,\text{ and }
  \hat\phi_1(t),\ldots,\hat\phi_M(t)$ are obtained as the eigenvalues
  and eigenvectors, respectively, from an eigendecomposition of the
  estimated covariance matrix.
\item The principal component scores are the best linear unbiased
  prediction (BLUP) estimates:
    \[
    \hbxi_i=\diag(\hat{\bnu})\hat\bPhi(\vect_i)^T\{\hat{G}(\vect_i,\vect_i)
    +\hat\sigma_x^2\matI_{N_i}\}^{-1}\{\widetilde{\vecx}_i-\hat{\bmu}_x(\vect_i)\},
    \]
    $\hbxi_i=(\hat\xi_{i1},\ldots,\hat\xi_{iM})^T$, and
    $\hat\bPhi(\vect_i)=[\hat\bphi_1(\vect_i):\cdots:\hat\bphi_M(\vect_i)]$,
    where $\hat{\bphi}_j(\vect_i)$ and
    $\hbmu_x(\vect_i)$ denotes the vector of evaluations of the
    $j$th estimated eigenfunction and estimated mean function,
    respectively, at the timepoints $\vect_i,\ i=1,\ldots,N.$
\end{enumerate}
The parameters $M$, $\mu(t),\ \nu_1,\ldots,\nu_M,\text{ and }
\phi_1(t),\ldots,\phi_M(t)$ are fixed at these initial estimates for
our MCMC and VB algorithms.  This is as done in
\textcite{goldsmith2011functional}, though they do update
$\nu_1,\dots,\nu_M$.  For ease of notation, we suppress the
``hat''/circumflex for these parameters when developing our algorithms
in later sections.  The principal component scores as well as the
measurement error variance are updated by both algorithms, and we will
demonstrate that our methods can be used to accurately estimate more
principal components beyond the first two.  This procedure is also
used in our numerical experiments when, for comparison, we also
estimate FGAM using the two-step approach mentioned in the
introduction.
\section{Penalized Spline Smoothing For FGAM}\label{mmform}
\subsection{Review of Tensor Product Splines}
\linelabel{ae:4}We next discuss our representation for the bivariate
surface
$F(\cdot,\cdot)$ in \eqref{fgam}.  We choose to use penalized splines
which are very popular tools for applied regression modelling.
Book-length treatments of penalized splines are provided by
\textcite{ruppert2003semiparametric,wood2006generalized}.  We use
tensor product splines to approximate
$F(x,t)$.  In this approach, the bivariate function
$F$ is constructed from univariate splines in $x$ and
$t$ as follows.  First, considering a univariate function, $f$ of
$x$, we may represent
$f(x)$ using splines as
$f(x)=\sum_{j=1}^{K_x}\gamma_jB^X_j(x)$, where the
$\gamma_j$ are spline coefficients to be estimated from the data and
$\{B^X_1(x),\ldots,B^X_{K_x}(x)\}$ is a prespecified
$K_x$-dimensional spline basis over the possible values of
$x$.  We add dependence on
$t$ by considering the spline coefficients to be functions of
$t$, which we again approximate using splines; i.e.\
$\gamma_j(t)=\sum_{k=1}^{K_t}\theta_{jk}B_k^T(t)$;
$j=1,\ldots,K_t$; where
$\{B^T_1(t),\ldots,B^T_{K_t}(t)\}$ is a spline basis for the
$t$-axis.  Combining the equations for $f(x)$ and the
$\gamma_j(t)$'s we have the following tensor product spline
representation for $F(x,t)$
\autocite[e.g.,][]{wood2006low}
\begin{equation}\label{tpsform}
F(x,t)=\sum_{j=1}^{K_x}\sum_{k=1}^{K_t}\theta_{jk}B_j^X(x)B_k^T(t).
\end{equation}

We use B-splines for the basis functions, which are popular because of
their good numerical properties and for their computational
convenience.  An introduction to univariate and tensor product
B-splines is provided in \textcite[Ch.~1,2]{dierckx1995curve}.  To fit
a tensor product spline model, one must specify a polynomial degree
for the univariate spline bases; the number of knots for the spline
bases, $K_x$ and
$K_t$, as well as their location; penalty parameters, $d_x$ and
$d_t$; and smoothing parameters, $\lambda_x$ and
$\lambda_t$ which control the trade-off between fitting the data
(minimizing mean square error) and complexity of
$F(\cdot,\cdot)$.  While this may seem like a lot of parameters to
specify, in practice assuming that $K_x$ and
$K_t$ are chosen to be large enough, the most important factor
determining the fit of the spline is the choice of smoothing
parameters \autocite{ruppert2002selecting}, and the knots are almost
always specified to be equally-spaced distances apart.  The degree of
the spline bases is typically chosen to be three, i.e.\ cubic splines
and the penalty parameters, which we discuss in more detail shortly,
are both chosen to be two, so that the surface is shrunk towards a
plane with increasing $\lambda_x$ and
$\lambda_t$ \autocite{eilers1996flexible}.  The cubic B-spline pairs,
$B^X_j(x)B^T_k(t)$, look like overlapping ``humps'' or standard
bivariate normal densities.

Plugging \eqref{tpsform} into \eqref{fgam} we obtain
\begin{align*}
E(Y_i\mid X_i) &= \eta_{0i} + \int F\{x_i(t),t\}\,dt\approx \eta_{0i} + \sum^{K_x}_{j=1} \sum^{K_t}_{k=1}  \int B^\mcX_j\{x_i(t)\}
          B^\mcT_k(t)\theta_{j,k}\,dt
\end{align*}
The integral above must be approximated via quadrature.  We specify a
grid of time points $\vect=(t_1,\ldots,t_T)^T$ where the integral is
to be evaluated and define a vector of quadrature weights,
$\vecL=(L_1,\ldots,L_T)^T$.  The estimated trajectories and the
B-spline bases are evaluated at $\vect$, which results in the vectors
$\vecx_i=\{x_i(t_1),\dots,x_i(t_T)\}^T$; i=1,\ldots,N;
$\vecB^\mathcal{X}_{j,i} = [B^\mathcal{X}_{j}\{x_i(t_1)\}$,
$\dots, B^\mathcal{X}_j\{x_i(t_T)\}]\tr$; and
$\vecB^\mathcal{T}_k = \{B^\mathcal{T}_k(t_1), \dots,
B^\mathcal{T}_k(t_T)\}\tr$.  For the $i$th subject and $(j,k)$ basis
function pair, we have
$\int B^X_j\{x_i(t)\}B^T_k(t)\,dt\approx
\vecL^T(\vecB^\mcX\odot\vecB^\mcX)$, where $\odot$ denotes
element-wise multiplication.  We then arrive at the following
approximation to \eqref{fgam} which we use for the rest of the paper
\begin{equation}
  E(Y_i\mid X_i) = \eta_{0i} + \int F\{x_i(t),t\}\,dt\approx \eta_{0i} + \sum^{K_x}_{j=1} \sum^{K_t}_{k=1}\vecL\tr (\vecB^\mcX_{j,i} \odot \vecB^{\mcT}_{k})\theta_{j,k}= \eta_{0i} + \sum^{K_x}_{j=1} \sum^{K_t}_{k=1} Z_{j,k,i} \theta_{j,k}.
\end{equation}
%
Recalling our estimate for the trajectories from the previous section,
for ease of notation, we will write $\bmu_x(\vect)$ and $\bPhi(\vect)$
as $\bmu_x$ and $\bPhi$, respectively, and only specify the grid of
evaluation points if it differs from $\vect$.  In the calculations
that follow, we frequently work with the $T\times K_xK_t$ matrix of
B-spline products evaluated at the grid points, $\vect$, and estimated
trajectories, $\vecx_i=\bmu_x+\bPhi\bxi_i$: \beq\label{Bmat}
\matB_{\xi_i}=\left[\{B_1^X(\bmu_x+\bPhi\bxi_i)\cdots
  B_{K_x}^X(\bmu_x+\bPhi\bxi_i)\}\otimes\mathbf{1}^T_{K_t}\right]\odot\left[\mathbf{1}_{K_x}^T\otimes\{B_1^T(\vect)\cdots
  B_{K_t}^T(\vect)\}\right],\quad i=1,\ldots,N, \eeq where $\otimes$
denotes the Kronecker product.  This matrix is always multiplied on
the left by the vector of quadrature weights, $\vecL$, so we also
define $\bxit\equiv\vecL^T\Bxi$.  Note that
$\bxit=(Z_{1,1,i},\dots,Z_{K_x,K_t,i})^T$ is the $i$th row of the
matrix $\matZ$ from \textcite{McLean2012functional}.
\subsection{Formulation as a Mixed Model}
The mixed model formulation of penalized splines is now well-known and
widely-used, see e.g., for a review.  The FGAM looks superficially
like a bivariate smoothing problem, but it is more challenging since
we do not observe $F(x,t)$ (with error) for pairs $(x,t)$ but instead
we observe only the integral of $F\{X(t),t\}$ with respect to $t$.
Nonetheless, some ideas from bivariate smoothing are applicable to
FGAM.  As in \textcite{McLean2012functional}, we start with a
bivariate spline model for $F(\cdot,\cdot)$ based on P-splines
\autocite{eilers1996flexible,marx2005multidimensional}.  We take a
more general approach than the Bayesian P-splines of
\textcite{lang2004bayesian}, which performed isotropic smoothing via a
first-order Gaussian random walk prior for the bivariate components in
their additive model.

Frequently in the penalized spline literature it is assumed
$\lambda_x=\lambda_t$ to simplify estimation.  Here however, because
$X(t)$ and $t$ having differing scales, it is not appropriate to
assume apriori that the amount of smoothing for $F(x,t)$ should be the
same in both arguments.  Though we may scale $x$ and $t$ to lie in the
unit square, this would still not result in a scale-invariant tensor
product smooth \autocite{wood2012straightforward}.  The necessitated
anisotropic roughness penalty associated with the spline coefficients,
$\btheta=(\theta_{11}, \dots, \theta_{1,K_t},\theta_{2,1}\dots,
\theta_{K_x,K_t})^T$, requires considerable more care than the
univariate smoothing necessary for the Bayesian FLM in
\textcite{goldsmith2011penalized}, the isotropic penalty used in
\textcite{mueller2013continuously}, or the penalized structured
additive regression literature
\autocite[e.g.,][]{fahrmeir2004penalized}.

\textcite{wahba1983bayesian} first made the connection between spline
smoothing and Bayesian modeling, showing that the usual (frequentist)
estimator for a cubic smoothing spline was equivalent to placing a
particular improper Gaussian prior on the spline coefficients.  The
penalization used in \textcite{McLean2012functional} is equivalent to
imposing the following prior on the spline coefficients
\begin{align*}
 p(\btheta\mid \lambda_x, \lambda_t) \propto \exp\left(-\frac{1}{2}\btheta\tr
 \matP_\btheta(\lambda_x, \lambda_t) \btheta \right),
\end{align*}
with
$\matP_\btheta(\lambda_x, \lambda_t) = \lambda_x \matP_x + \lambda_t
\matP_t$, with $\matP_x = \matD_x\tr\matD_x \otimes \matI_{K_t}$,
$\matP_t = \matI_{K_x} \otimes \matD_t\tr\matD_t$.  $\matI_p$ is the
identity matrix of dimension $p$, $\matD_t$ and $\matD_x$ are
difference operator matrices of the prespecified degrees, $d_x$ and
$d_t$, respectively.  This penalty structure leads to a partially
improper Gaussian prior since $\matP_\btheta(\lambda_x, \lambda_t)$ is
rank deficient: $\matD_x\tr\matD_x$ has rank $K_x - d_x$,
$\matD_t\tr\matD_t$ has rank $K_t - d_t$, so that
$\matP_\btheta(\lambda_x, \lambda_t)$ has rank $K_x K_t- d_x d_t$
\autocite[Section 4.4]{horn1994topics}. To avoid numerical instability
associated with inversion of numerically rank-deficient matrices when
sampling from the full conditional of $\btheta$ and the appearance of
the zero determinant of $\matP_\btheta(\lambda_x, \lambda_t)$ in the
full conditionals of $\lambda_x$ and $\lambda_t$, we aim for a simpler
representation of the function by employing the mixed model
representation of tensor product splines used in \textcite[Section
6]{currie2006generalized}.  The idea is to simultaneously diagonalize
the marginal penalties for $x$ and $t$.  This results in a diagonal
penalty structure which is efficient for computations and easy to
interpret.

More precisely, we split the function $F(x,t)$ into an unpenalized
part parameterizing functions from the nullspace of the penalty (i.e.,
associated with a diffuse Gaussian prior on the coefficients) and a
penalized part (associated with a non-diffuse Gaussian prior on the
coefficients).  We begin by rewriting the vector of function
evaluations for subject $i$ as
$F( \vecx_i, \vect) = \sum^{K_x}_j \sum^{K_t}_k
(\vecB^\mathcal{X}_{j,i} \odot \vecB^\mathcal{T}_{k}) \theta_{j,k} =
\matB_{\xi_i}\btheta.$ We take the spectral decompositions of the
marginal penalties, i.e.,
\[
\matD_x\tr\matD_x = \matV_x\matS_x\matV_x\tr,\; \matD_t\tr\matD_t = \matV_t\matS_t\matV_t\tr,
 \]
 where both $\matV_x$ and $\matV_t$ are orthogonal matrices and
 $\matS_x$ and $\matS_t$ are diagonal.  We define
 $\widetilde{\matV}_x\text{ and }\widetilde{\matV}_t$ to be the
 matrices of eigenvectors associated with zero eigenvalues, which have
 dimension $K_x \times d_x\text{ and }K_t \times d_t,$ respectively.
 The basis functions for the unpenalized part of the tensor product
 spline can then be defined as
 $\matB_{i,0} = \matB_{\xi_i}(\widetilde{\matV}_t \otimes
 \widetilde{\matV}_x)$,

 For the basis for the penalized part of the tensor product spline,
 $\matB_{i,p}$, we first define
 $\matS_{t,x} = (\matI_{K_t} \otimes \matS_x) + (\matS_t \otimes
 \matI_{K_x})$, a matrix that has all combinations of sums of the
 eigenvalues on the diagonal, and form $\widetilde{\matS}_{t,x}$,
 which is $\matS_{t,x}$ without the zero entries on the diagonal
 corresponding to $\matB_{i,0}$. This can be written as
 $\widetilde{\matS}_{t,x} = \matU\tr\matS_{t,x}\matU$, where $\matU$
 is a $K_x K_t \times (K_x K_t - d_xd_t)$ orthogonal matrix
 constructed by removing $d_xd_t$ columns from $\matI_{K_x K_t}$.  We
 thus have
\[
  \matB_{i,p} = \matB_{\xi_i} (\matV_t \otimes \matV_x) \matU
  \widetilde{\matS}_{t,x}^{-1/2}, \text{ so that }
  \matB_{\xi_i}\btheta = \matB_{i,0} \bbeta + \matB_{i,p}\bdelta
\]
  or, for clearer exposition,
\[
  \matB_{\xi_i}\btheta = (\matB_{\xi_i}\matT)
  (\matT^{-1}\btheta)\text{ with } \matT
  =[\matT_0:\matT_p]=\bmat{2}(\widetilde{\matV}_t \otimes
  \widetilde{\matV}_x):(\matV_t \otimes \matV_x) \matU
  \widetilde{\matS}_{t,x}^{-1/2} \emat,
  \]
 and
 $\matT^{-1} = \bmat{2} (\widetilde{\matV}_t \otimes
 \widetilde{\matV}_x) : (\matV_t \otimes \matV_x) \matU
 \widetilde{\matS}_{t,x}^{1/2} \emat\tr$.

 The penalty matrix $\matP_\btheta(\lambda_x,
 \lambda_t)$ of the reparameterized coefficient vector $(\bbeta\tr,
 \bdelta\tr)\tr =
 \matT^{-1}\btheta$ becomes $\widetilde{\matP}_\btheta(\lambda_x,
 \lambda_t)=$ $\matT\tr \matP_\btheta(\lambda_x, \lambda_t)
 \matT.$ Since $\matP_\btheta(\lambda_x, \lambda_t)\matT_0 =
 0$, only the lower right $(K_x K_t - d_xd_t)\times(K_x K_t -
 d_xd_t)$-quadrant of $\widetilde{\matP}_\btheta(\lambda_x,
 \lambda_t)$ is of interest.  Denoting this submatrix by
 $\widetilde{\matP}_\bdelta(\lambda_x,
 \lambda_t)$, our penalty is now given by the diagonal matrix
\[
  \widetilde{\matP}_\bdelta(\lambda_x, \lambda_t) = \lambda_t \bPsi_t
  + \lambda_x \bPsi_x;\quad \bPsi_t =
  \widetilde{\matS}_{t,x}^{-1/2}\matU\tr(\matS_t \otimes
  \matI_{K_x})\matU \widetilde{\matS}_{t,x}^{-1/2};\quad\bPsi_x =
  \matI_{K_x K_t - d_x d_t} - \bPsi_t;
\]
see \textcite{currie2006generalized}.

Recalling that $\bxit=\vecL^T\matB_{\xi_i}$ with $\matB_{\xi_i}$ given
by \eqref{Bmat}, we can now
write 
\[
\int F(X_i(t),t)\,dt\approx \bxit\btheta=\vecL\tr \matB_{\xi_i}\matT\binom{\bbeta}{\bdelta}=\vecL\tr \matB_{\xi_i} \matT_0 \bbeta + \vecL\tr \matB_{\xi_i} \matT_p\bdelta = \vecL\tr \matB_{i,0}\bbeta + \vecL\tr \matB_{i,p}\bdelta.
\]

We use diffuse inverse gamma (IG) priors for the variance components and our full model is given by
\begin{align*}
  Y_i&\sim N(\eta_{0i}+\vecL\tr \matB_{i,0}\bbeta + \vecL\tr \matB_{i,p}\bdelta,\sigma^2);\quad \sigma^2\sim \text{IG}(a_e,b_e);\\
  \tilde{\vecx}_i(\vect_i)&\sim N(\mu_x(\vect_i)+\bPhi(\vect_i)\bxi_{i},\sigma^2_x\matI_{n_i});\quad \sigma_x^2\sim\text{IG}(a_x,b_x);\\
  \xi_{im}&\sim N(0,\nu_m);\quad m=1,\ldots,M;\numberthis\label{fullmodel}\\
  \bdelta&\sim N\left(0,[\lambda_t \bPsi_t + \lambda_x \bPsi_x]^{-1}\right);\quad \lambda_x,\lambda_t\sim \text{Gamma}(a_l,b_l);\\
  \bbeta&\sim N(0,\sigma_\beta^2\matI_{d_xd_t});\quad \eta_{0i}\sim N(0,\sigma_\eta^2);\quad i=1,\ldots,N
\end{align*}
\section{An MCMC algorithm for fitting FGAM}
We now describe an MCMC algorithm for fitting FGAM.  We will use a
Metropolis-within-Gibbs sampler.  The conjugate priors used for the
spline coefficients and the variance components (excluding the
smoothing parameters) in our hierarchical model allow for closed-form
expressions for those parameters' full conditional distributions.
Since their derivations are quite standard, we omit the details until
Appendix A and focus in this section on the more complicated updates
for the smoothing parameters and principal component scores.

To understand what is being updated and in what order, we start by
providing pseudocode outlining the updates made by our MCMC algorithm
to sample the posterior of model \eqref{fullmodel}.  Details of how
the updates are done will be provided
subsequently.\linelabel{editor:1}This pseudocode also applies to our
variational Bayes algorithm developed in the next section; the change
being that instead of parameters being updated by randomly drawing
from a distribution that converges in the limit to the true posterior
distribution (subject to regularity conditions), they are
deterministic updates of hyperparameters and moments of optimal
densities.  The pseudocode is given in Algorithm \ref{pseudocode}.

\begin{algorithm}
\caption{Pseudocode for fitting FGAM given by \eqref{fullmodel}}
\label{pseudocode}
\begin{algorithmic}[1]
  \State Obtain initial estimates, $\vecx$, for the trajectories using
  the method from Section \ref{PACE}.  \State Specify penalties and
  bases for $F(x,t)$. Obtain decomposition from Section \ref{mmform}.
  \State Initialize other parameters.
  \Repeat
  \For{$i = 1 \to N$}
  \State Update principal component scores, $\bxi_i$.
  \State Update $\vecx_i$.
  \State Update $\matB_{i,p}$.
  \EndFor
  \For{$i = 1 \to N$}
  \State Update terms involving scalar covariates, $\eta_{0i}$.
  \EndFor
  \State Update unpenalized spline coefficients, $\bbeta$.
  \State Update penalized spline coefficients, $\bdelta$.
  \State Update smoothing parameters, $\lambda_x,\ \lambda_t$.
  \State Update measurement error variance, $\sigma_x^2$.
  \State Update response error variance, $\sigma^2$.
  \Until{Maximum number of iterations reached \emph{OR} [for VB] convergence criteria met.}
\end{algorithmic}
\end{algorithm}

The updates for $\lambda_x$ and $\lambda_t$ require special attention
because of the non-conjugality of their full conditional
distributions.  To see this, we have
\begin{align*}
 p(\lambda_x \mid \text{rest}) &=p(\lambda_x\mid\lambda_t,\bdelta)\propto p(\bdelta\mid\lambda_x,\lambda_t)p(\lambda_x)\propto\lvert\lambda_x\bPsi_x +  \lambda_t \bPsi_t\rvert^{1/2}(\lambda_x)^{a_l + 1} \exp\{-(b_l + \frac{1}{2} \bdelta \tr
\bPsi_x \bdelta ) \lambda_x\} \\
 & \propto \lvert\lambda_x \bPsi_x +  \lambda_t
\bPsi_t\rvert^{1/2} \Gamma(\text{shape}=a_l + 2, \text{scale}= \{b_l + \frac{1}{2} \bdelta \tr\bPsi_x \bdelta \}^{-1})\equiv f_{\lambda_x}(\lambda_x),\numberthis\label{spfc}
\end{align*}
where ``rest'' is used to denote all parameters and data in the model
besides $\lambda_x$.  The derivation is analogous for $\lambda_t$.  We
do not obtain a closed-form expression for these full conditionals
because of the determinant in \eqref{spfc}.  We overcome this
difficulty by using slice sampling \autocite{neal2003slice}.  Slice
sampling is a method for efficiently sampling from nonstandard
distributions such as \eqref{spfc} by alternatingly sampling from the
vertical region under $f_{\lambda_x}(x)$ and then sampling from the
horizonal region under the density at the location of the vertical
sample.  \textcite[Section 8]{neal2003slice} demonstrated that slice
sampling can be more efficient than Metropolis methods for fitting
Bayesian hierarchical models.

In our implementation, given an initial value, $\lambda_0$, and defining $g(x):=\log[f_{\lambda_x}(x)]$, we obtain a draw $\lambda_1$ from $p(\lambda_x\mid\text{rest})$ as follows
\begin{enumerate}
\item Draw $u\sim \text{Unif}\{0,g(\lambda_0)\}$ which defines a "slice" $S:=\{x:u<g(x)\}$
\item Obtain an interval $[L,R]$ such that $S \subset[L,R]$ by
  starting with $[L_0,R_0]=[0,2]$ and expanding the interval until
  $[L,R]$ contains $S$
\item Draw $\lambda_1\sim \text{Unif}(L,R)$.  If $\lambda_1\not\in S$,
  shrink $[L,R]$ and draw $\lambda_1$ again until $\lambda_1\in S$,
\end{enumerate}
and analogously for
$\lambda_t$.  For further details including proof of convergence to
the proper posterior, see \textcite{neal2003slice}; his Figure 1 is
especially recommended for building intuition.

The second difficulty in developing our MCMC algorithm occurs when
updating the principal component scores.  This stems from the
likelihood being a nonlinear function of the scores (they appear as
arguments to B-spline basis functions).  We have
\begin{align*}
p(\bxi_i\mid\text{rest}) &\propto p(y_i\mid\eta_{0i},\bbeta,\bdelta,\bxi_i,\sigma^2)p(\tilde{\vecx}_i\mid\bxi_i,\sigma_x^2)p(\bxi_i)\\
&\propto\exp\left\{-\frac{1}{2 \sigma^2}\left[
    y_{\eta_0,i} -  \sum^{K_x}_j \sum^{K_t}_k \vecL \tr \left\{\vecB^\mathcal{X}_j\left(\bmu_x + \bPhi \bxi_{i}\right)\odot \vecB^\mathcal{T}_{k}(\vect)\right\}
          \theta_{j,k} \right]^2 \right\} \cdot\\
&\quad \cdot \exp\left[-\frac{\{\widetilde{\vecx}_{\mu,i} - \bPhi(\vect_{i})\bxi_{i}\}\tr\{\widetilde{\vecx}_{\mu,i} - \bPhi(\vect_{i})\bxi_{i}\}}{2 \sigma_x^2}\right] \cdot\exp\left\{-
\frac{\bxi_i \tr \diag(\bnu^{-1}) \bxi_i}{2}\right\}
\intertext{where $\widetilde{\vecx}_{\mu,i} = \tilde{\vecx}_{i} - \mu_x(\vect_i)$ and $y_{\eta_0, i} =  y_i - \eta_{0i}$, so that}
p(\bxi_i\mid\text{rest})&= \exp\left\{-\frac{1}{2\sigma^2}\left[ y_{\eta_0, i} -  \sum^{K_x}_j \sum^{K_t}_k  \theta_{j,k} \sum^T_t
     L_t B^\mathcal{X}_j\left\{\mu_x(t_t) + \bPhi(t_t)^T\bxi_{i}\right\} B^\mathcal{T}_{k}\{t_t\}\right]^2 \right\} \cdot\\
    &\quad \cdot N\left[\vecm_{\xi, i}=\matS_{\bxi, i} \bPhi(\vect_{i})^T\widetilde{\vecx}_{\mu,i},\; \matS_{\bxi, i} = \left\{\bPhi(\vect_{i})\tr\bPhi(\vect_{i})/\sigma_x^2 +
     \diag(\bnu^{-1})\right\}^{-1}
    \right]
\end{align*}

We update each $\bxi_i,\; i=1,\dots,n$ based on its full conditional,
with a proposal density for new values, $\bxi_i^\star$, based only on
the trajectories and a Metropolis-Hastings (M-H) acceptance correction
to account for the intractable part of the full conditional involving
the likelihood of $\vecy$.

Specifically, the proposal distribution is 
\begin{align*}
q_1(\bxi_i,\bxi_i^\star) = N\left[\vecm_{\xi, i}=\matS_{\xi, i} \bPhi(\vect_{i})^T\widetilde{\vecx}_{\mu,i}/\sigma_x^2,\; \matS_{\xi, i} = \left\{\bPhi(\vect_{i})\tr\bPhi(\vect_{i})/\sigma_x^2 + \diag(\bnu^{-1})\right\}^{-1}\right],
\end{align*}
 so that 
$q_1(\bxi_i,\bxi_i^\star) = q_1(\bxi_i^\star)$ independent of the current state. The acceptance
probability $\alpha(\bxi_i,\bxi_i^\star)$ is then given by
\begin{align*}
 &\frac{q_1(\bxi_i^\star,\bxi_i) p(\bxi_i^\star\mid\cdot)}
 {q_1(\bxi_i,\bxi_i^\star)p(\bxi_i\mid\cdot)} \wedge 1= \frac{ \exp\left\{-\frac{1}{2 \sigma^2}\left[
    y_{\eta_0,i} -  \sum^{K_x}_{j=1} \sum^{K_t}_{k=1} \vecL \tr \left\{\vecB^\mathcal{X}_j\left(\bmu_x + \bPhi \bxi_{i}^\star\right)\odot \vecB^\mathcal{T}_{k}(\vect)\right\}
          \theta_{j,k} \right]^2 \right\}}{\exp\left\{-\frac{1}{2 \sigma^2}\left[
    y_{\eta_0,i} -  \sum^{K_x}_{j=1} \sum^{K_t}_{k=1} \vecL \tr \left\{\vecB^\mathcal{X}_j\left(\bmu_x + \bPhi \bxi_{i}\right)\odot \vecB^\mathcal{T}_{k}(\vect)\right\}
          \theta_{j,k} \right]^2 \right\}}\wedge 1
\end{align*}
because the ratio of proposal distributions cancels with the ratio of
the tractable parts of the full conditionals.


As we will see in our numerical studies, the implausible trajectories
that occasionally result from an FPCA occur much less frequently in
our MCMC approach.  This is because the proposals of extreme PC scores
are likely to be rejected by our M-H step since they seem even more
implausible when considered along with the response and current
estimates of the regression coefficients in the acceptance
probability.

The formula for the full model posterior can be found in Appendix A.
\section{A Variational Bayes Approach}\label{secvb}
In this section we develop a variational Bayes algorithm for fitting
the FGAM.  We begin with a quick review of variational approximations.
\subsection{Review of Variational Bayes}
Our notation in this section closely follows that of
\textcite{goldsmith2011functional}.  We define
$\mu_{q(\theta)}\equiv E_q(\theta)=\int \theta_0
q_\theta(\theta_0)\,d\theta_0$ and
$\sigma^2_{q(\theta)}\equiv
\text{Var}_q(\theta)=\int\{\theta_0-E_q(\theta)\}^2q_\theta(\theta_0)\,d\theta_0$
for scalar parameters, and analogously define $\mu_{q(\btheta)}$ and
$\Sigma_{q(\btheta)}$ for vector parameters. We will give a brief
overview of the main ideas of VB, and refer the reader to
\textcite[Chapter 10]{bishop2006pattern} or
\textcite{jaakkola2000bayesian} for further details.  Given observed
data $\vecy$ and a collection of parameters $\btheta$, the goal of
variational Bayes is to find a simplified density $q(\btheta)$ that
approximates the desired posterior $p(\btheta\mid\vecy)$ as closely as
possible according to Kullback-Leibler (KL) divergence.  The
derivation of a variational Bayes algorithm relies on the result from
\textcite{kullback1951information} that for an arbitrary density,
$q(\btheta)$, the marginal likelihood, $p(\vecy)$, satisfies
$p(\vecy)\geq \underline{p}(\vecy;q):=\exp\left[\int
  q(\btheta)\log\left\{{p(\vecy;\btheta)}/{q(\btheta)}\right\}d\btheta\right]$,
with equality if and only if $q(\btheta)=p(\btheta\mid\vecy)$.

While other simplifications, for example that the density of interest,
$q(\btheta)$, is parametric, are sometimes used for variational
approximations, variational Bayes uses the assumption that a posterior
density can be factorized as $q(\btheta)=\prod_{p=1}^P q_p(\btheta_p)$
for some partition $\{\btheta_1,\ldots,\btheta_P\}$ of $\btheta$.
Assuming this factorization for $q$ and using the above result on KL
divergence, it is easy to show \autocite[see
e.g.,][]{ormerod2010explaining} that $\underline{p}(\vecy;q)$ is
maximized when $q_p$ is chosen to be \beq\label{optdensity}
q^*_p(\btheta_p)\propto\exp\left[E_{-\btheta_p}\{\log
  p(\vecy,\btheta)\}\right]\propto\exp\left[E_{-\btheta_p}\{\log
  p(\btheta_p\mid\text{rest})\}\right];\quad p=1,\ldots,P; \eeq where
$E_{-\btheta_p}[\cdot]$ denotes expectation w.r.t.\ all model
parameters excluding $\btheta_p$.  We thus have a deterministic
algorithm where one full iteration updates each component $\btheta_p$
sequentially using $q^*_p(\btheta_p)$.  The algorithm terminates when
the change in $\underline{p}(\vecy;q)$ becomes sufficiently small.
Notice that the density in \eqref{optdensity} is precisely the full
conditional from Gibbs sampling, and the optimal density is tractable
when the full conditional is conjugate.

Helpful tools for deriving VB algorithms are directed acyclic graphs
(DAGs) and Markov blankets.  A Markov blanket is the set of all child,
parent, and co-parent nodes of a particular node in a DAG.  Examples
can be found in \textcite[Chapter 8]{bishop2006pattern}.  Calculating
the densities in \eqref{optdensity} is made much simpler because of
the result that
$p(\btheta_p\mid\text{rest})=p(\btheta_p\mid\text{Markov blanket of
}\btheta_p)$.
\subsection{Fitting FGAM Using Variational Bayes}
Our VB algorithm for fitting FGAM follows the same general steps used
by our MCMC approach and given in Algorithm~\ref{pseudocode}.  As with
MCMC, updates for the spline coefficients and variance components
(smoothing parameters excluded) follow from standard calculations, so
we leave them to \ref{app_optden}.  The non-standard updates of the
principal component scores and smoothing parameters are discussed
below.

\begin{figure}
\centering
\includegraphics[width=4in]{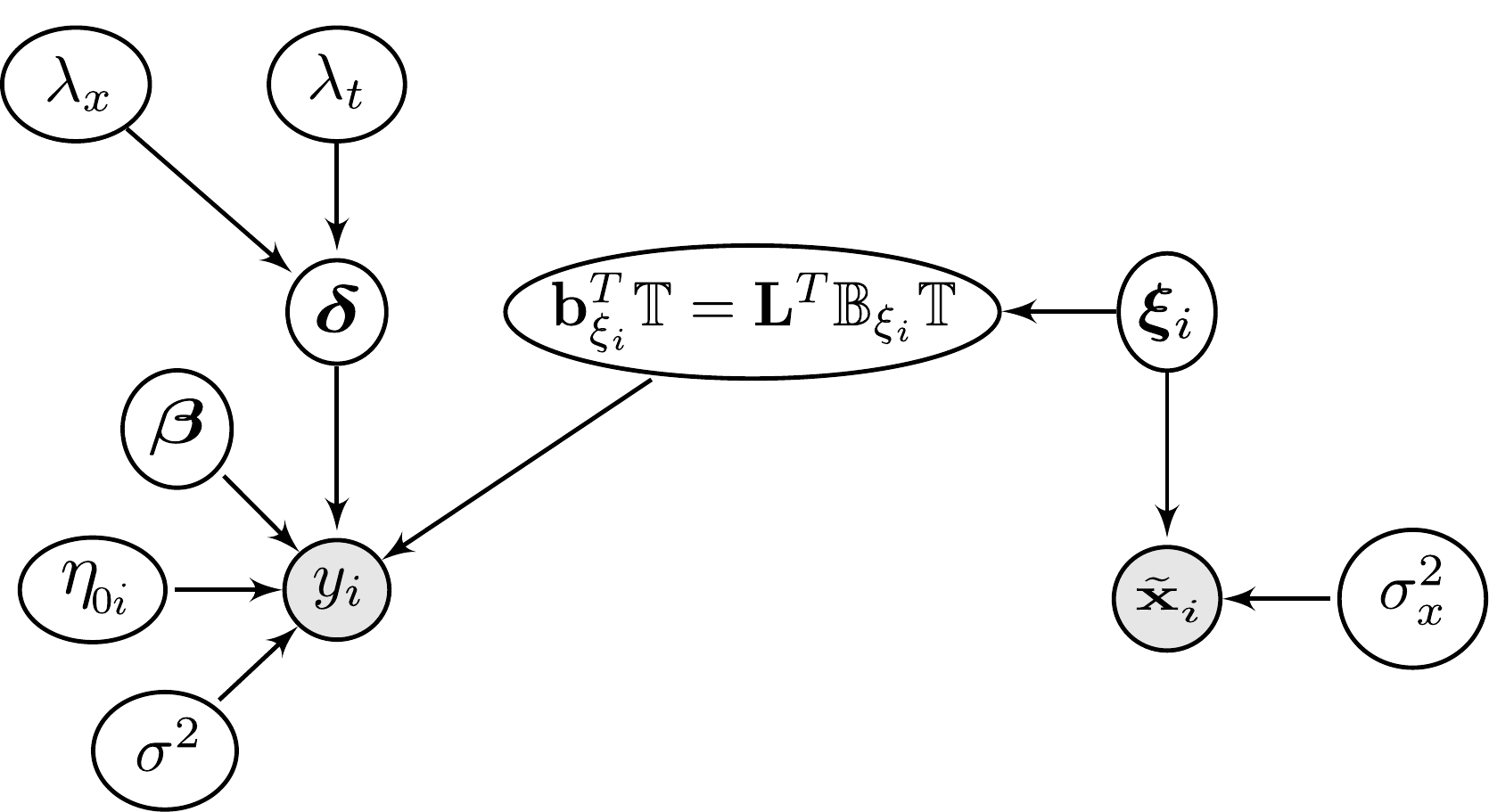}
\caption{Directed Acyclic Graph for FGAM.  Shaded vertices denote known quantities.  The parameters $\{\nu_m\},\ \{\phi_m\},\ M,\text{ and }\mu_x$ are omitted since they are not updated by the VB algorithm.}\label{dagfig}
\end{figure}

Using $\bTheta$ to denote all unknown parameters in our model
\eqref{fullmodel}, we assume the posterior
$p(\bTheta\mid\vecy,\tilde{\vecx})$ admits the factorization
$p(\bTheta\mid\vecy,\tilde{\vecx})=q(\bbeta)q(\bdelta)q(\lambda_x)q(\lambda_t)q(\sigma^2)q(\sigma_x)\prod_{i=1}^Nq(\bxi_i)q(\eta_{0i})$.
The DAG for FGAM is shown in Figure~\ref{dagfig}.

For the optimal density for $\lambda_x$, we have from \eqref{optdensity}
\begin{align*}
q^*&(\lambda_x)\propto \exp\left[\E_{-\lambda_x}\{\log p(\lambda_x\mid\text{rest})\}\right]\\
&=\exp\left[\E_{-\lambda_x}\left\{\frac{1}{2}\log\lvert\lambda_x \bPsi_x +  \lambda_t \bPsi_t\rvert-\frac{1}{2} \bdelta^T (\lambda_x \bPsi_x)
\bdelta +(a_l + 1)\log(\lambda_x) -b_l \lambda_x\right\}\right]\\
&\approx \exp\left[\frac{1}{2}\log\lvert\lambda_x \bPsi_x +  \Mqlt \bPsi_t\rvert-\frac{\lambda_x}{2} \left\{\text{tr}(\bPsi_x\Sqdelta)+ \Mqdelta^T\bPsi_x\Mqdelta\right\}+(a_l + 1)\log(\lambda_x) -b_l \lambda_x\right]\\
&=\lvert\lambda_x \bPsi_x +  \Mqlt \bPsi_t\rvert^{1/2}\exp\left[ -b_l \lambda_x-\frac{\lambda_x}{2} \left\{\text{tr}(\bPsi_x\Sqdelta)+ \Mqdelta^T\bPsi_x\Mqdelta\right\}\right]\lambda_x^{a_l + 1}\equiv\tql{x},\numberthis\label{qlambda}
\end{align*}
where the approximation comes from plugging in $\Mqlt$ for $\lambda_t$
to avoid taking an expectation of the determinant term over
$\lambda_t$.  Notice
$c_{q(\lambda_x)}\equiv\int_0^\infty\tq{x}{x}\,dx$ has the form
$c_{q(\lambda_x)}=\int_0^\infty x^{a_l+1}e^{-x}f(x)\,dx$ which can be
approximated by generalized Gauss-Laguerre quadrature.
\linelabel{ref2:3}Using this type of quadrature for variational Bayes
is discussed in \textcite{wand2011mean} and is implemented in
\texttt{R} in the package \texttt{statmod} \autocite{statmod}, and we
use it to determine a grid of $G$ points, $\vecg$, and quadrature
weights, $\vecL_g$.  Our approximations are then given by
$c_{q(\lambda_x)}\approx \vecL_g^T\widetilde{q}_{\lambda_x}(\vecg)$
and
$\Mqlx\approx\{\vecL_g^T\widetilde{q}_{\lambda_x}(\vecg)\}^{-1}\vecL_g^T\left\{\vecg\odot\widetilde{q}_{\lambda_x}(\vecg)\right\}.$

Due to the exponential term in \eqref{qlambda}, moderate to large
values of $\lambda_x$ result in $\tilde{q}_{\lambda_x}(\lambda_x)$
being evaluated to be zero, unless care is taken during the
computation to avoid underflow.  One strategy for avoiding loss of
precision is as follows.  Define $\ell_{\lambda_x}(x)=\log\tq{x}{x}$
and $m_{\lambda_x}=\max_\vecg \ell_{\lambda_x}(\vecg)$, then
$c_{q(\lambda_x)}\approx
\exp(m_{\lambda_x})\vecL_g^T\exp\{\ell_{\lambda_x}(\vecg)-m_{\lambda_x}\}$.
The term $\exp(m_{\lambda_x})$ is in both the numerator and the
denominator of $\Mqlx$ and thus drops out in that calculation.  Taking
the logarithm of the determinant in $\tql{x}$ is not a problem because
$\bPsi_x$ and $\bPsi_t$ are
diagonal.

For updating the principal component scores in our VB algorithm,
recall the form of the full conditional
\begin{align*}
&p(\bxi_i\mid\text{rest})\propto p(y_i\mid\eta_{0i},\bbeta,\bdelta,\bxi_i,\sigma^2)p(\tilde{\vecx}_i\mid\bxi_i,\sigma^2_x)p(\bxi_i)\\
&\propto \exp\left\{-\frac{1}{2\sigma^2}(y_i-\eta_{0i}-\bxit\btheta)^2\right\}\exp\left\{-\frac{1}{2\sigma_x^2}\lVert\tilde{\vecx}_i-\bmu_x(\vect_i)-\bPhi(\vect_i)\bxi_i\rVert^2_2\right\}
\exp\left\{-\frac{1}{2}\bxi_i^T\diag(\bnu^{-1})\bxi_i\right\},
\end{align*}
where as before $\bxit=\vecL^T\matB_{\xi_i}$ with $\matB_{\xi_i}$ given by \eqref{Bmat}.  We have,
\begin{align*}
\text{E}_{-\bxi_i}&\left\{-\frac{1}{2\sigma^2}(y_i-\eta_{0i}-\bxit\btheta)^2\right\}=-\frac{\Mqisigma}{2}\E_{-\bxi_i}\left[\{y_i- \mu_{q(\eta_{0i})}-\E_{-\bxi_i}(\bxit\btheta)\}^2\right]\\
&-\frac{1}{2}\Mqisigma\sigma^2_{q(\eta_{0i})}-\frac{\Mqisigma}{2}\E_{-\bxi_i}\left\{(\bxit\btheta-\E_{-\bxi_i}(\bxit\btheta))^2\right\}\\
&=-\frac{\Mqisigma}{2}\left[(y_i-\mu_{q(\eta_{0i})} -\bxit\Mqtheta)^2+\sigma^2_{q(\eta_{0i})}+\E_{-\bxi_i}\left\{(\btheta-\Mqtheta)^T\vecbxi\bxit(\btheta-\Mqtheta)\right\}\right]\\
&=-\frac{\Mqisigma}{2}\left\{(y_i- \mu_{q(\eta_{0i})}-\bxit\Mqtheta)^2+\sigma^2_{q(\eta_{0i})}+\trace(\vecbxi\bxit\Sqtheta) \right\}
\end{align*}
Therefore,
\begin{align*}
q^*(\bxi_i)&\propto\exp\left[-\frac{\Mqisigma}{2}\left\{(y_i- \mu_{q(\eta_{0i})}-\bxit\Mqtheta)^2+\sigma^2_{q(\eta_{0i})}+\trace(\vecbxi\bxit\Sqtheta)
\right\}\right.\\
&\qquad\left.-\frac{\Mqisigmax}{2}\lVert\tilde{\vecx}_i-\bmu_x(\vect_i)-\bPhi(\vect_i)\bxi_i\rVert^2_2-\frac{1}{2}\bxi_i^T\diag(\bnu^{-1})\bxi_i\right]\\
&\propto\exp\left[\Mqisigma\{y_i- \mu_{q(\eta_{0i})}\}\bxit\Mqtheta-\frac{\Mqisigma}{2}\left\{(\bxit\Mqtheta)^2+\bxit\Sqtheta\vecbxi\right\}\right.\\
&\qquad\left.+\Mqisigmax\{\tilde{\vecx}_i-\bmu_x(\vect_i)\}^T\bPhi(\vect_i)\bxi_i-\frac{1}{2}\bxi_i^T\left\{\Mqisigmax\bPhi^T(\vect_i)\bPhi(\vect_i)+\diag(\bnu^{-1})\right\}\bxi_i\right]\equiv q(\bxi_i).
\end{align*}
Since this does not have the form of a standard, known density, we
will employ a Laplace approximation.  The use of Laplace
approximations for variational inference with nonconjugate models was
also explored in \textcite{wang2013variational}.  This is given
by
\begin{equation}\label{laplace}
q^*(\bxi_i)=N(\bxi_{i,0},\bLambda_i^{-1})\quad\text{where}\quad \bLambda_i=\left.-\mcD_{\bxi_i^T}\mcD_{\bxi_i}\log q(\bxi_i)\right|_{\bxi_i=\bxi_{i,0}},
\end{equation}
with $\mcD_\veca[\cdot]$ denoting differentiation w.r.t.\ the vector
$\veca$ and $\bxi_{i,0}$ denoting the mode of $q^*(\bxi_i)$, which is
found by a numerical optimization routine.  The formula for
$\bLambda_i$ is given in \ref{app_optden}.  We expect the Laplace
approximation to perform well in high sparsity settings because the
Gaussian prior becomes the dominant part of the posterior in these
situations.

To construct our algorithm, we also require the expectation of
$\vecbxi$ and the expectation of its outer product with respect to
$\bxi_i$.  To do this we use second-order Taylor expansions about
$\bxi_{i,0}$.  These derivations are also left to \ref{app_optden}.
Our log-likelihood lower bound, which is used for monitoring
convergence of our algorithm, is derived in \ref{app_llbound} and the
full variational Bayes algorithm is given in \ref{app_vbalg} as
Algorithm \ref{VBalgorithm}.
\section{Simulation Study}
\begin{figure}
\centering
\includegraphics[width=\linewidth]{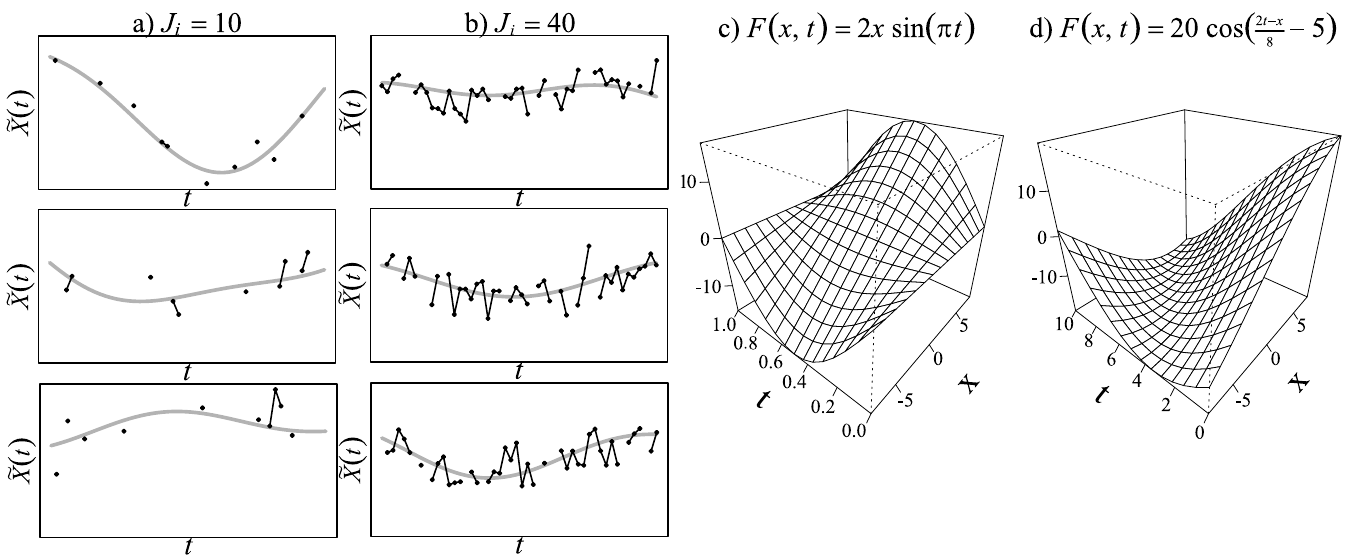}
\caption{\small Plots a) and b) show three observed functional predictors for varying levels of sparsity when $\sigma_x=1$.  The true trajectories are also plotted in grey.  Plot c) shows the surface $F(x,t)=2x\sin(\pi t)$ and plot d) the surface $F(x,t)=20\cos\left(-\frac{x}{8}+\frac{t}{4}-5\right)$.\label{simdata}}
\end{figure}

We now conduct a simulation study to compare the efficacy of our
proposed approaches.  We fit each model to 100 simulated data sets.
The true functional covariates are given by
$X(t)=\sum_{j=1}^{4}\xi_j\phi_{j}(t),$ with $\xi_j\sim N(0,8j^{-2})$
and
$\{\phi_1(t),\ldots,\phi_4(t)\}=\{\sin(\pi
t/\lvert\mcT\rvert),\cos(\pi t/\lvert\mcT\rvert),\sin(2\pi
t/\lvert\mcT\rvert),\cos(2\pi t/\lvert\mcT\rvert)\}$, with
$\rvert\mcT\lvert$ denoting the measure of the interval $\mcT$.  To
examine how our model performs with both sparse and dense but
irregularly observed data, we generate observed covariates by randomly
selecting $J_i=10$ or $J_i=40$ points for each subject from a grid of
50 equally-space points used to generate the true response.  We
consider three different levels of the measurement error variance,
$\sigma_x^2=0,1,\text{ and }4$.  The response error variance is taken
to be $\sigma^2=1$.  We examine two different possibilities for the
regression surface $F(x,t)$.  First, a case where the FLM is the true
model, $F(x,t)=2x\sin(\pi t)$, with $\mcT=[0,1]$; and next, a case
where the FLM does not hold,
$F(x,t)=20\cos\left(-\frac{x}{8}+\frac{t}{4}-5\right)$, with
$\mcT=[0,10]$. A sampling of some generated curves including
measurement error for both levels of sparsity as well as plots of both
true surfaces can be found in Figure~\ref{simdata}.

For our comparison we consider seven different methods for fitting
FLMs and FGAMs: 1) a baseline/oracle FGAM fit by the
\textcite{McLean2012functional}\ approach when the fully observed
curves without measurement error are known (trueX), 2) FGAM fit by
\textcite{McLean2012functional} with fixed trajectories estimated
using the procedure outlined in Section~\ref{PACE} (PACE), 3) FGAM fit
using variational Bayes on the sparse, noisy curves (VB), 4) FGAM fit
using MCMC and the sparse, noisy curves (MCMC), 5) as in 4) except
initial values are supplied by the VB fit (VB-MCMC), 6) FLM fit using
penalized splines with trajectories obtained from the
Section~\ref{PACE} procedure (FLM-PACE), and 7) FLM fit to the fully
observed curves without measurement error (FLMtrueX).  Each method
used cubic B-splines and second-order difference penalties.  The
\textcite{McLean2012functional} implementation of FGAM is fit using
their code which is available in the package \texttt{refund}
\autocite{refund} in \texttt{R} \autocite{r2012}.  Smoothing
parameters are chosen by generalized cross validation (GCV) using the
package \texttt{mgcv} \autocite{wood2011fast}, which is also used to
estimate the FLMs.  MCMC runs one chain for 10,000 iterations after a
burn-in of 1000, whereas VB-MCMC uses only 1000 iterations after a
burn-in of 500.  Each method uses and, if applicable, estimates
exactly the true number of non-zero components $M=4$.  For each
simulated data set, we use two thirds of the 100 observations to fit
the models and the other one third for prediction.

\begin{figure}
\centering
%
\executeiffilenewer{Sim_RIMSEX_F.svg}{Sim_RIMSEX_F.pdf}%
{inkscape -z -D --file=Sim_RIMSEX_F.svg %
--export-pdf=Sim_RIMSEX_F.pdf --export-latex}%
\resizebox{!}{.3\textheight}{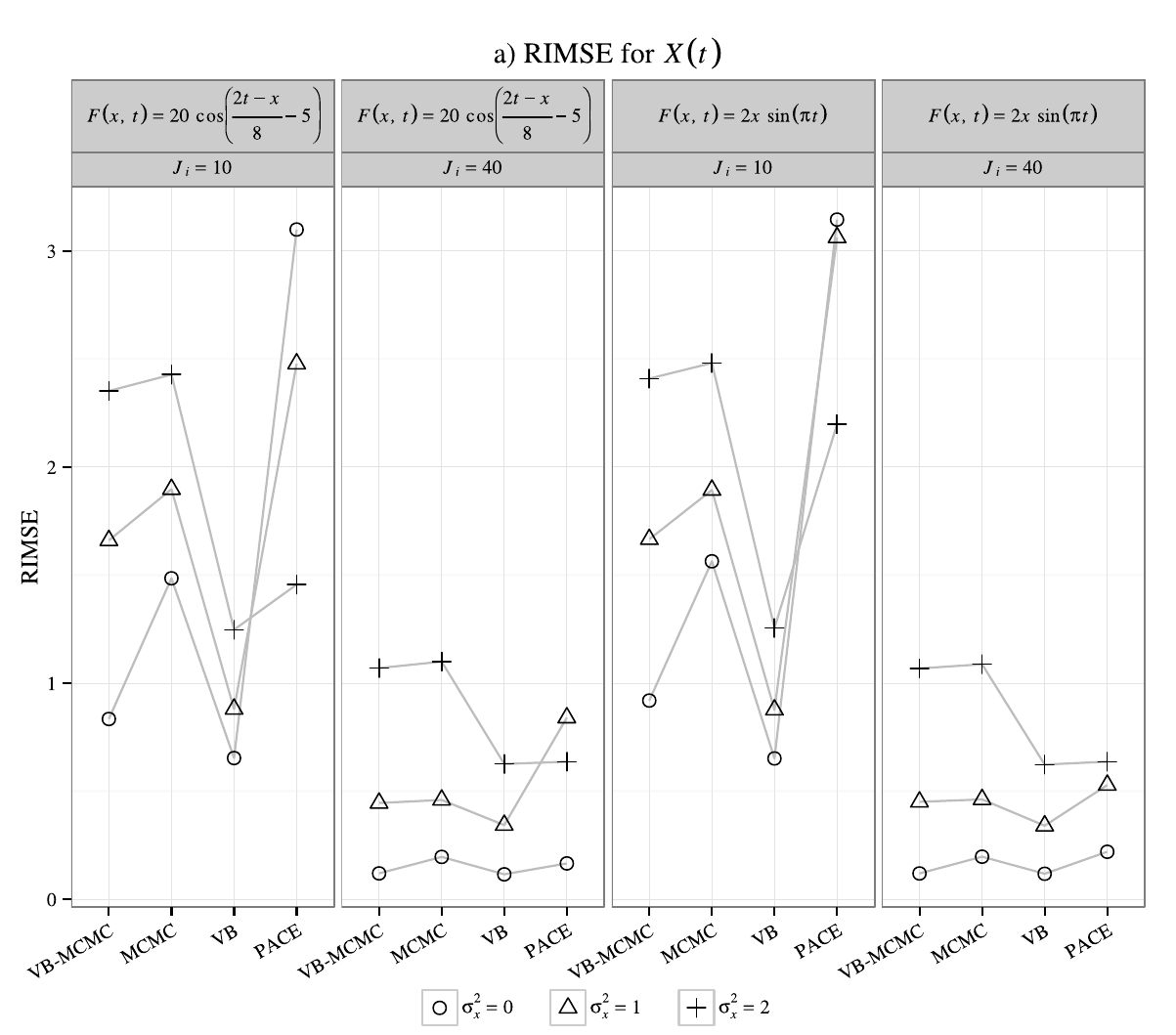}

\executeiffilenewer{Sim_RIMSEFin_F.svg}{Sim_RIMSEFin_F.pdf}%
{inkscape -z -D --file=Sim_RIMSEFin_F.svg %
--export-pdf=Sim_RIMSEFin_F.pdf --export-latex}%
\resizebox{!}{.3\textheight}{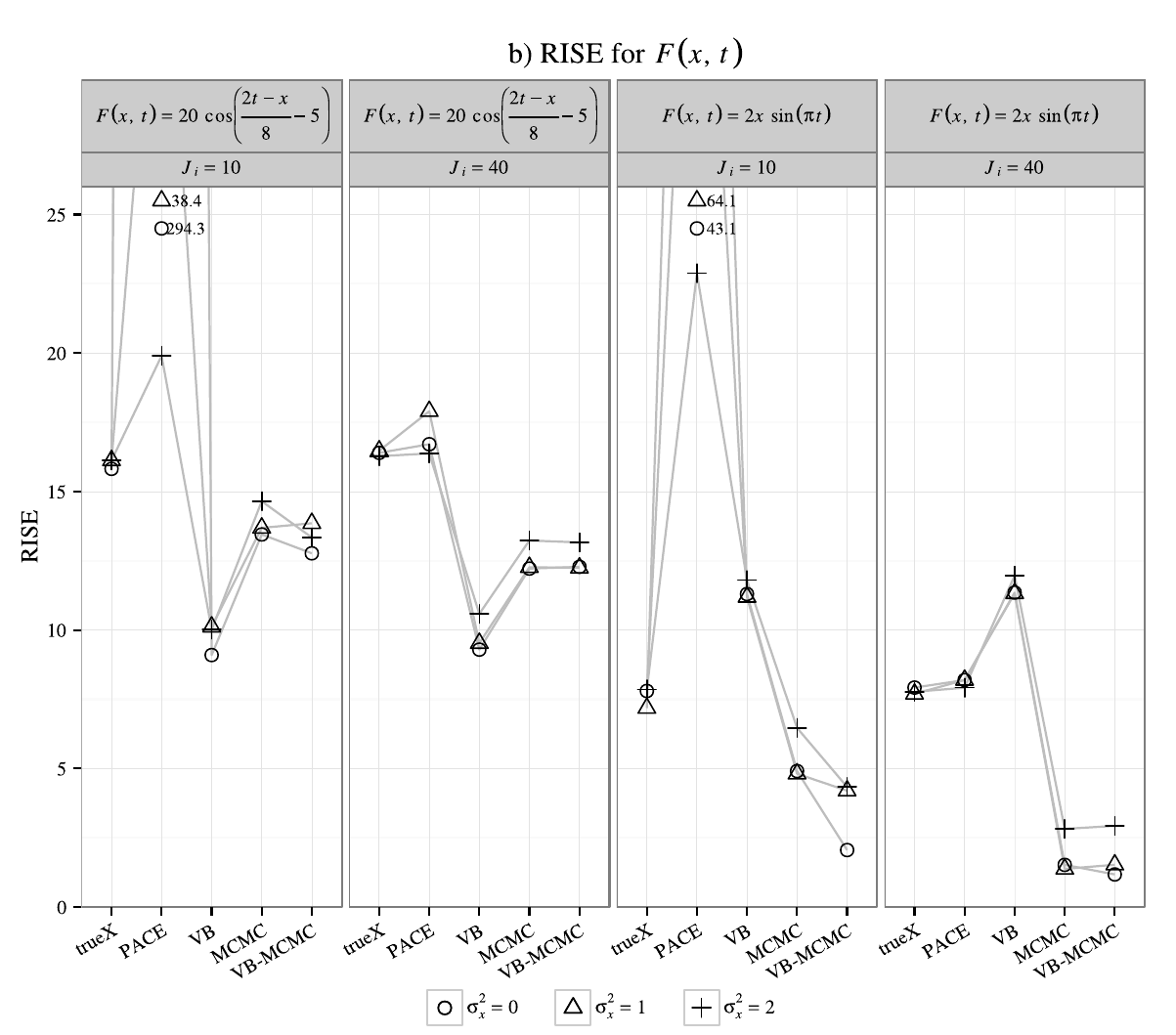}

\caption{\small a) Median RMISE over 100 simulations for two levels of
  sparsity and different values for the measurement error variance for
  recovering in-sample trajectories, $X(t)$. b) Median RISE for
  predicting the true surface, $F(x,t)$.  b) includes trueX which is
  not relevant for a).  Values that do not fall within the y-axis
  limits are individually labeled.\label{XFrmse}}
\end{figure}

We first compare how well PACE, VB, MCMC, and VB-MCMC do at estimating
the functional covariates.  The median over simulations of the
in-sample root mean integrated square error,
${\text{RMISE-X}}^2=N^{-1}\sum_{i=1}^{67}\int_{\mcT}\{X_i(t)-\hat{X}_i(t)\}^2\,dt$,
for each scenario and method is reported in Figure~\ref{XFrmse} a).
We see that the PACE method does not perform well in the sparse data
scenarios $(J_i=10)$.  One reason for this is that it does not account
for the variability from imputing the principal component scores.  An
additional reason is difficulties in estimating a covariance matrix
for the functional predictors.  The estimate is often singular or
near-singular and this causes numerical problems when attempting to
estimate all four non-zero principal component scores using the method
presented in Section~\ref{PACE}.  Our Bayesian algorithms do not
suffer from this problem even when starting from poorly conditioned
initial estimates from our PACE implementation.  We see that VB
performs quite well at recovering the trajectories, even in the
$J_i=10$ scenarios.  MCMC performs slightly worse than VB here.
Further investigation showed that MCMC on average slightly
overestimated $\sigma_x^2$ which made it less accurate for in-sample
recovery, but that this added variance made for more accurate
prediction of trajectories out-of-sample.  \linelabel{editor:2a}The
observed acceptance rates for the independent Metropolis-Hastings step
used to update the principal component scores were consistently above
0.9 for all scenarios indicating that our proposal distribution
performed well for this data.

Now turning to estimation of the true surface $F(x,t)$, we report the
median root integrated square error,
RISE-F$^2=\int_\mcX\int_\mcT\left\{F(x,t)-\hat{F}(x,t)\right\}^2\,dt\,dx$,
in Figure~\ref{XFrmse} b).  We evaluate the RISE only at $(x,t)$
values that are inside the convex hull defined by the observed
trajectories for that sample to avoid regions of the plane where there
are no data.  We again observe performance from the PACE method to be
poor in the sparse settings.  Interestingly, the MCMC and MCMC-VB
approaches have lower ISE than the trueX method.  We suspect this is
due to the MCMC algorithm on average choosing larger smoothing
parameters which are closer to the optimal values for smoothing the
surface than those chosen by GCV for the trueX fits.  Due to the
additional smoothing performed by the integration in \eqref{fgam}, the
optimal amount of smoothing for estimating the response and for
estimating the surface are different \autocite{cai2006prediction}.
Also noteworthy is the substantial difference between VB and MCMC
depending on the true regression surface.  This again seems to be due
to differences in how the smoothing parameters are chosen.

\begin{figure}
\centering
%
\executeiffilenewer{Sim_RMSEynew_F.svg}{Sim_RMSEynew_F.pdf}%
{inkscape -z -D --file=Sim_RMSEynew_F.svg %
--export-pdf=Sim_RMSEynew_F.pdf --export-latex}%
\resizebox{!}{.3\textheight}{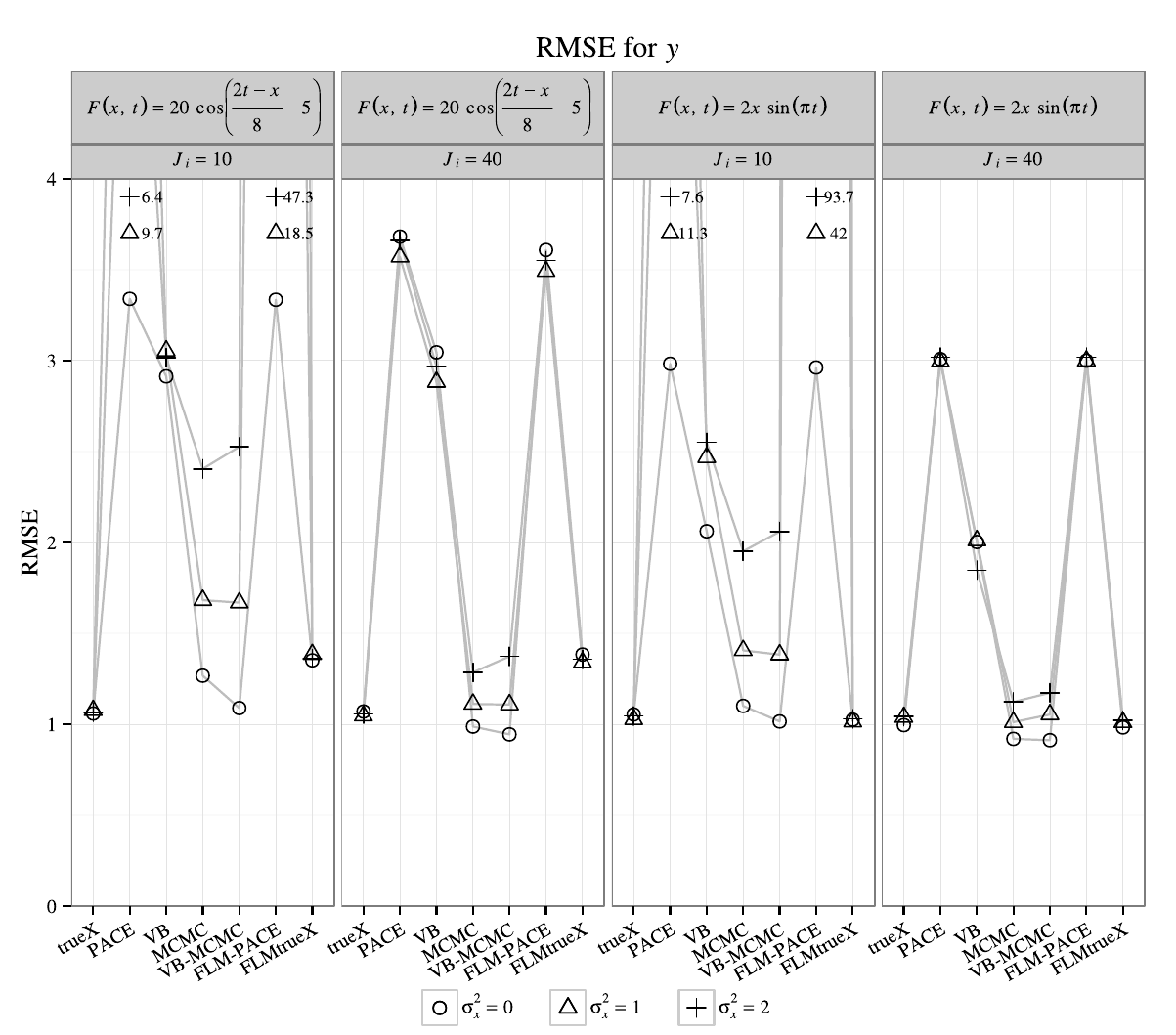}

\caption{\small Median RMSE over 100 simulations for out-of-sample
  predictions of the response, $Y$, for two levels of sparsity and
  different values for the measurement error variance.  Values that do
  not fall within the y-axis limits are individually
  labeled\label{Yrmse}}
\end{figure}

Finally, results for root mean square error (RMSE) for predicting the
out-of-sample response,
$\text{RMSE-Y}^2=\frac{1}{33}\sum_{i=68}^{100}(Y_i-\hat{Y}_i)^2$, can
be found in Figure~\ref{Yrmse}.  We see that the performance of MCMC
matches and even sometimes outperforms the oracle trueX method that
knows the entire trajectories.  Overall, we recommend the combination
of VB for initial estimates followed by MCMC as it appears to be best
or close to best in nearly all scenarios.  The total elapsed time for
estimating FGAM on one data set averaged over all simulations and
scenarios was 43.3 seconds for VB, 732.0 seconds for MCMC, and 153.5
seconds for VB-MCMC.
\section{Analysis of Auction Data}\label{auction}
In this section we fit our proposed models to auction data from the
online auction website eBay and attempt to forecast closing auction
price.  The data set contains the time and amount of every bid for 155
seven-day auctions of Palm M515 Personal Digital Assistants (PDA) that
took place between March and May, 2003.  Each auction is
"standardized" to start at time 0.  This data was previously analyzed
using functional data methods in a series of work by W.\ Jank, G.\
Shmueli and coauthors
\autocite[e.g.,][]{jank2006functional,wang2008explaining}.  The PACE
methodology introduced in Section~\ref{PACE} was used to analyze this
data set in \textcite{liu2008functional}.  Typically, each auction
consists of three clearly discernible parts: an initial period with
some bidding, a middle period with very few bids, and a final period
of rapid bidding as the auction finishes
\autocite{wang2008explaining}.  This sparsity and irregularity in the
observed bid data means that the usual methods of function data
analysis are not appropriate.

Our raw data is actually the maximum amount the bidder is willing to
pay for the item, often called the willing-to-pay (WTP) value.  To
recover the current item price from the WTP values, we must use the
table available at
\url{http://pages.ebay.com/help/buy/bid-increments.html}.  When a new
WTP value is entered that is more than any previous WTP value, the new
price is determined by incrementing the current price in an amount
given by this table.  A new bidder must enter an amount at least as
large as this new price plus the increment given by the table.  We
assume there is an underlying smooth price process that we attempt to
recover with our proposed approaches.

We use the logarithm of the ratio of successive prices during the
first six days of the auction to predict the logarithm of the closing
price on the final day.  Hourly prices are used so that we are trying
to recover $6\times24=144$ prices for each auction.  When an auction
has multiple bids in the same hour, we take the average of the prices
corresponding to those bids as the observed price for that hour.  As
in \textcite{liu2008functional}, we set any negative values for the
log-price ratio equal to zero, which can occur because initial
log-price at time $0$ is taken to be zero.  To show the usefulness of
our MCMC and VB methods, we fit the FGAM and FLM using the trajectory
of observed log-price ratios,
log$\{\tilde x_i(t_{i,j})/\tilde x_i(t_{i,j-1})\}$, for the first six
days in order to predict the logarithm of the final selling price at
the end of the seventh day.  We emphasize that no information on the
prices from the final day of the auction are included in the
functional predictor so that we have a true measure of forecasting
accuracy.

We randomly partition the data into training and test sets with two
thirds of the samples used for training and one third for testing.  We
compute the root mean square error (RMSE) for predicting the logarithm
of the closing price for the test data set after fitting each model to
the training data.  This is repeated for 25 different splits into test
and training sets.  For comparison, we also considered the simple
two-step approach of using PACE to recover the functional predictors
and then using these estimates to fit FLMs and FGAMs in
\texttt{refund} as in the fully-observed predictor case from
\textcite{McLean2012functional}.  For the FGAM methods, ten basis
functions were used for both
axes.

\begin{figure}[!ht]
\centering
\includegraphics[height=3.5in]{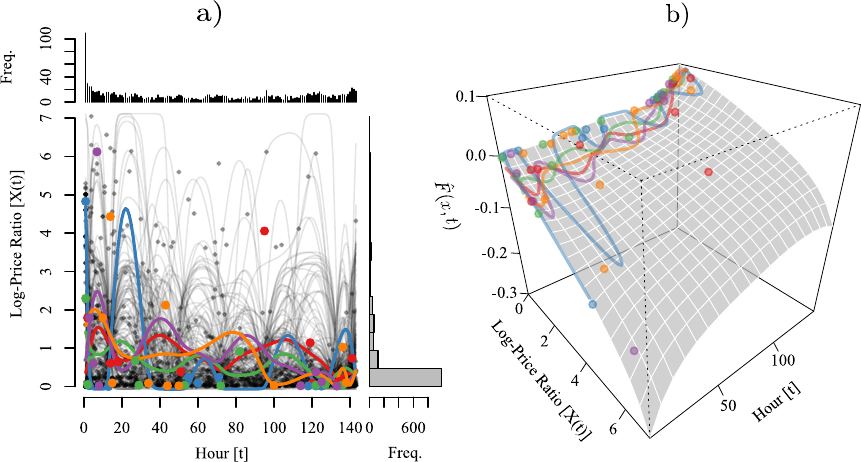}
\caption{\small a) All estimated trajectories from use of our MCMC
  algorithm on the auction data with points representing observed
  data.  Five trajectories are highlighted and also plotted in b).
  Also included are two histograms showing which covariate values
  occur with the highest frequency (on the right) and the frequency of
  bids for each hour of the auction (on top).  b) Shows the estimated
  surface $\hat{F}(x,t)$ from fitting FGAM to the auction data using
  MCMC.  The overlayed points and curves are the same as
  a)\label{mcmcfit}}
\end{figure}

The surface estimated by our MCMC algorithm fit to the entire data set
is displayed in Figure~\ref{mcmcfit}~b) along with the observed and
estimated log-price ratios for five randomly chosen auctions.
Figure~\ref{mcmcfit}~a) plots all estimated trajectories and
additionally histograms showing the frequencies of observations for
both $X(t)$ and $t$; notice from the histogram on the right part of
the plot that the majority of the data is grouped at very low
log-price ratios.  In b) we see that large values of the log-price
ratio in the early hours of the auction result in a lower predicted
value for the closing price and that smaller ratios later towards the
end of the sixth day of the auction result in higher predicted closing
price.  Nonlinearities in the log-price component of the estimated
surface suggest that an FLM may not be flexible enough for this data
set.  There appears to be some undersmoothing of the functional
predictors in Figure~\ref{mcmcfit}~a).  \textcite{cai2006prediction}
showed that for optimal prediction in the FLM, the coefficient
function should be undersmoothed because of the additional smoothing
performed by the integral in the regression function.  We conjecture
that some degree of undersmoothing of the functional predictors is
desirable for our forecasting problem when estimating
\eqref{fullmodel} for similar
reasons.

The median out-of-sample RMSE over 25 partitions of the data is
reported in Table \ref{auctionres} along with standard deviations.  We
can see that our Bayesian approach for fitting FGAM offers the best
performance in this case, with both FGAM-MCMC and FGAM-VB offering
much improved performance over the methods that only use PACE followed
by estimation of FGAM in \texttt{refund}.  Both methods that simply
used PACE and then assumed fully observed data had very poor
performance for some of the splits when the imputed trajectories were
especially
bad.  
\begin{table}
\centering
    \begin{tabular}{|c|c|c|c|}
      \hline
       FLM-PACE & FGAM-PACE &  FGAM-MCMC & FGAM-VB\\
      \hline
       0.5917(1.3093) & 4.913(0.4322) & 0.0914(0.0052) & 0.0905(0.0037) \\
      \hline
    \end{tabular}
    \caption{Median RMSE (with standard deviation in parentheses) for
      out of sample predictions of log-final selling price for 25
      random splits of the auction data\label{auctionres}}
\end{table}
\section{Conclusion}
We have proposed two algorithms for fitting a nonlinear regression
model for scalar on function regression when the functional predictor
is sparsely observed with measurement error.  After first expressing
the FGAM as a linear mixed model with missing data, we then took a
Bayesian hierarchical modeling approach and fit our model using a
Metropolis-within-Gibbs sampler.  Our MCMC algorithm was able to
provide useful inferences in difficult situations where initial
estimates provided by standard FPCA methods were quite poor due to
rank deficiency in the estimated covariance matrix.

Additionally, we developed a variational Bayesian algorithm for
fitting FGAM which can be used to quickly obtain approximate parameter
estimates.  We demonstrated the usefulness of our approach using
simulated data and an application to a longitudinal data set involving
online auctions.  We developed a Laplace approximation that accurately
approximated the intractable optimal density for the principal
component scores.  We also demonstrated the usefulness of using the
estimates from our VB algorithm as inputs to the MCMC algorithm to
obtain faster convergence.

An alternative way to account for uncertainty in the imputed
trajectories would be using the bootstrap approach of
\textcite{goldsmith2012corrected}.  We did implement this method, but
due to space concerns, we have not included it in this work.  In our
experiments, this approach did not perform as well as our Bayesian
algorithms and was slower than the combined VB-MCMC approach.

\linelabel{editor:4a}An interesting area for future work brought up by
a referee is that of using the estimates from a VB algorithm in a more
principled way to achieve faster convergence of an MCMC algorithm than
the simple approach considered in this work of using the VB estimates
as starting values for the MCMC.  Naively using a variational
approximate distribution as a proposal density or as a prior
distribution in a MCMC algorithm could be problematic due to the
tendency for variational Bayes to underestimate the true variance.
\textcite{de2001variational} demonstrated some algorithms combining
MFVB and MCMC that attempted to deal with this issue.
Additional areas for future work include investigating coverage for
credible bands provided by our variational Bayes algorithm and
comparing with credible intervals from MCMC.  Typically, credible
bands derived from variational Bayes procedures suffer from
undercoverage.  Bootstrapping the estimates from our variational Bayes
algorithm may be a promising way around this issue
\autocite{goldsmith2011functional}.  Another promising approach for
correcting covariance estimates from MFVB was recently proposed in
\textcite{giordano2015linear}.  We are also working on extensions to
the case of functional responses and binary responses.
\section*{Acknowledgements}
Much of this work was completed when Mathew McLean was a PhD student
at Cornell University supported by an NSERC PGS-D award.  We thank
Wolfgang Jank for providing the auction data.
\appendix
\renewcommand{\thesection}{Appendix~\Alph{section}}
\section{Derivation of Full Conditional Distributions}\label{app_fc}
In this appendix we derive the full conditional distributions for the
variance components and spline coefficients in \eqref{fullmodel} and
also give the full posterior distribution.
\subsection*{Variance parameters}
We begin by defining the $N\times d_xd_t$ matrix $\matZ_0$ whose $i$th
row is given by
$\vecZ_{0,i}^T=\vecL^T\matB_{\xi_i}\matT_0=\bxit\matT_0$ and the
$N\times (K_xK_t-d_xd_t)$ matrix $\matZ_p$ with $i$th row given by
$\vecZ_{p,i}^T=\bxit\matT_p$.  We also define
$\vecy_{\eta_0}=(y_1-\eta_{0,1},\ldots,y_N-\eta_{0,N})^T$,
$\boldeta_1=\matZ_0\bbeta+\matZ_p\bdelta$ with $i$th component
$\eta_{1,i} = \bxit\{\matT (\bbeta\tr, \bdelta
\tr)\tr\}=\bxit\matT_0\bbeta+\bxit\matT_p\bdelta,$, and
$\bXi=[\bxi_1:\cdots:\bxi_N]^T$, we have
\begin{align*}
p(\sigma^2\mid\cdot) &\propto p(\vecy\mid\eta_{0,1},\ldots,\eta_{0,N},\bbeta,\bdelta,\sigma^2,\bXi)p(\sigma^2)\\
&\propto(\sigma^2)^{-N/2 - a_s -1}\exp\left\{-\frac{b_s + \frac{1}{2} \left(\vecy_{\eta_0} - \boldeta_1\right)\tr \left(\vecy_{\eta_0} - \boldeta_1\right)}{\sigma^2} \right\} \\
\text{so that } \sigma^2\mid\cdot &\sim
\text{IG}\left(a = N/2 +
a_s, b = b_s + \frac{1}{2} \{\vecy_{\eta_0} - \boldeta_1\}\tr\{\vecy_{\eta_0} - \boldeta_1\}\right)\\ \intertext{Similarly,} \sigma_x^2\mid\cdot &\sim  \text{IG}\left(a= \sum^N_i n_i /2 + a_x, b= b_x +
\frac{1}{2} \sum^N_i  \sum_j^{n_i} \left\{\widetilde x_{ij} - \mu_x(t_{ij}) - \sum^M_m
\phi_{m}(t_{ij})\xi_{im}\right\}^2\right).\\
\end{align*}
\subsection*{Spline coefficients $\beta, \bdelta$}
\begin{align*}
p(\bbeta, \bdelta \mid\cdot) &\propto p(\bbeta)p(\bdelta\mid\lambda_x,\lambda_t)p(\lambda_x)p(\lambda_t)\\
&\propto\exp\left\{ -\frac{ (\vecy - \boldeta_{0} - \matZ_0\bbeta - \matZ_p \bdelta )\tr(\vecy - \boldeta_{0}
-  \matZ_0\bbeta - \matZ_p \bdelta ) }{2 \sigma^2} \right\}
\exp\left\{ -\frac{1}{2} \bdelta^T(\lambda_x \bPsi_x +  \lambda_t \bPsi_t) \bdelta \right\}
 \intertext{i.e.}
\bdelta \mid\cdot &\sim N(\vecm_b, \matS_b) \text{ with }\\
\matS_b &= (\matZ_p\tr\matZ_p/\sigma^2 + \lambda_x \bPsi_x +  \lambda_t\bPsi_t)^{-1}, \\
 \vecm_b &= \matS_b \matZ_p \tr \left(\vecy_{\eta_0} - \matZ_0 \bbeta\right)/\sigma^2;\\
\bbeta\mid\cdot &\sim N(\vecm_\beta, \matS_\beta) \text{ with }\\
\matS_\beta &= (\matZ_0\tr\matZ_0/\sigma^2)^{-1}, \\
\vecm_\beta &= \matS_\beta \matZ_0 \tr \left(\vecy_{\eta_0} - \matZ_p \bdelta \right)/\sigma^2.
\end{align*}

The full posterior distribution is given by
\begin{align*}
p(\bbeta, \bdelta, &  \sigma^2, \lambda_x, \lambda_t, \bXi, \sigma^2_x \mid \vecy, \tilde{\vecx}, \eta_{0,1},\ldots,\eta_{0,N}, \bmu_x, \bPhi, \bnu)\propto \\
&\propto (\sigma^2)^{-N/2} \exp\left[-\frac{1}{2 \sigma^2}\sum_i^N\left\{
    y_{\eta_0,i} -  \sum^{K_x}_{j=1} \sum^{K_t}_{k=1} \vecL \tr \left[\vecB^\mathcal{X}_j\left(\bmu_x + \bPhi\bxi_{i}\right)\cdot \vecB^\mathcal{T}_{k}(\vect)\right]
          [\matT (\bbeta\tr, \bdelta \tr)\tr]_{j,k} \right\}^2 \right] \cdot
          \nonumber\\ &\quad\cdot (\sigma_x^2)^{-\sum^N_i n_i / 2} \exp\left[-\frac{1}{2 \sigma_x^2}\sum_i^N \lvert\tilde{\vecx}_{i} - \mu_x(\vect_{i}) - \bPhi(\vect_{i})\bxi_{i}\rvert^2_2\right] \cdot \nonumber\\
 &\quad\cdot \lvert\lambda_x \bPsi_x +  \lambda_t \bPsi_t\rvert^{1/2}
\exp\left( -\frac{1}{2} \bdelta \tr (\lambda_x \bPsi_x +  \lambda_t \bPsi_t)
\bdelta \right)\cdot \nonumber\\ &\quad\cdot \exp\left(-\frac{1}{2}\sum_i^N \bxi_i \tr \diag(\bnu^{-1}) \bxi_i\right)\cdot (\sigma^2)^{-a_s-1} \exp(-b_s/\sigma^2) \cdot (\sigma_x^2)^{-a_x-1} \exp(-b_x/\sigma_x^2) \cdot \nonumber\\
 &\quad\cdot (\lambda_x)^{a_l + 1} \exp(-b_l \lambda_x) (\lambda_t)^{a_l + 1}  \exp(-b_l \lambda_t), \nonumber\\
\end{align*}
where $[\matA]_{j,k}$ denotes the entry in the $j$th row and $k$th column of the matrix $\matA$.
\section{Derivation Of Optimal Proposal Densities}\label{app_optden}
In this section we derive the optimal densities, $q^*$, for parameters
that were given conjugate priors and give detailed calculations for
our Laplace approximation to the optimal density for the principal
component scores.  We use the notation and full conditionals from
Appendix A and often make use of the results that for
$\vecx\sim(\bmu,\bSigma),\
\E[\vecx^T\vecS\vecx]=\trace(\vecS\bSigma)+\bmu^T\vecS\bmu\text{ and
}\E[\vecx\vecx^T]=\E[\vecx]\E[\vecx]^T+\var[\vecx].$

We first discuss the updates for the offset terms, $\eta_{0i}$,
$i=1,\ldots,N$.  For simplicity, we assume that they can be expressed
as $\eta_{0i}=\vecu_i^T\boldeta_0$ or
${(\eta_{01},\ldots,\eta_{0N})^T=\matU\boldeta_0}$, where $\matU$ is
an $N\times p_0$ matrix with rows $\vecu_i^T$ containing, for e.g.,
scalar covariate observations for parametric terms, basis function
evaluations for nonparametric terms, or a leading column of ones for
an intercept.  Further generalizations are straightforward.  The
coefficient vector $\boldeta_0$ has prior density
$p(\boldeta_0)=N(\veczero,\sigma^2_{\boldeta_0}\matI_{p_0}),$ with
$\sigma^2_{\boldeta_0}$ large and fixed.  The full conditional is
given by
\begin{align*}
p(\boldeta_0\mid\text{rest})&\propto p(\vecy\mid\boldeta_0,\bbeta,\bdelta,\bXi,\sigma^2)p(\boldeta_0)
\propto\exp\left[ -\frac{ (\vecy - \matU\boldeta_{0} - \boldeta_1 )^T(\vecy - \matU\boldeta_{0} -  \boldeta_1) }{2 \sigma^2}-\frac{1}{\sigma^2_{\boldeta_0}}\boldeta_0^T\matI_{p_0}\boldeta_0 \right]\\
&\propto\exp\left\{-\frac{1}{2}\left[\boldeta_0^T\left(\frac{1}{\sigma^2}\matU^T\matU+\frac{1}{\sigma^2_{\boldeta_0}}\matI_{p_0}\right)\boldeta_0
-2\left((\vecy- \boldeta_1 )^T\matU/\sigma^2\right)\boldeta_0\right]\right\}.
\end{align*}

Thus,
\begin{align*}
q^*(\boldeta_0)&\propto\exp\left\{-\frac{1}{2}\E_{-\boldeta_0}\left[\boldeta_0^T\left(\frac{1}{\sigma^2}\matU^T\matU+\frac{1}{\sigma^2_{\boldeta_0}}\matI_{p_0}\right)\boldeta_0
-2\left((\vecy- \boldeta_1 )^T\matU/\sigma^2\right)\boldeta_0\right]\right\}\\
&\propto\exp\left\{-\frac{1}{2}\left[\boldeta_0^T\left(\Mqisigma\matU^T\matU+\frac{1}{\sigma^2_{\boldeta_0}}\matI_{p_0}\right)\boldeta_0
-2\left((\vecy - \Mqetao )^T\matU\Mqisigma\right)\boldeta_0\right]\right\},
\end{align*}
where $\Mqetao=\MqBxi\matT(\Mqbeta^T,\Mqdelta^T)^T$.  Denote the rows of the $N\times K_xK_t$ matrix, $\MqBxi$, by $\Mqbxit$.  By completing the square, we see $q^*(\boldeta_0)=N(\Mqetaz,\Sqetaz)$ where
$\Sqetaz=\left(\Mqisigma\matU^T\matU+\frac{1}{\sigma^2_{\boldeta_0}}\matI_{p_0}\right)^{-1}$ and $\Mqetaz=\Sqetaz\matU^T(\vecy - \Mqetao )\Mqisigma$.

Next, for $\bbeta$
\begin{align*}
p(\bbeta\mid\text{rest})&\propto p(\vecy\mid\boldeta_0,\bbeta,\bdelta,\bXi,\sigma^2)p(\bbeta)
\propto\exp\left[ -\frac{ (\vecy - \matU\boldeta_{0} - \boldeta_1 )^T(\vecy - \matU\boldeta_{0} -  \boldeta_1) }{2 \sigma^2}-\frac{1}{\sigma^2_{\bbeta}}\bbeta^T\matI_{d_xd_t}\bbeta \right]\\
&\propto\exp\left\{-\frac{1}{2}\left[\bbeta^T\left(\frac{1}{\sigma^2}\matZ_0^T\matZ_0+\frac{1}{\sigma^2_{\bbeta}}\matI_{d_xd_t}\right)\bbeta
-2\left((\vecy- \matU\boldeta_{0} - \matZ_p\bdelta)^T\matZ_0/\sigma^2\right)\bbeta\right]\right\}
\intertext{Thus,}
q^*(\bbeta)&\propto\exp\left\{-\frac{1}{2}\left[\bbeta^T\left(\Mqisigma\E_{-\bbeta}[\matZ_0^T\matZ_0]+\frac{1}{\sigma^2_{\bbeta}}\matI_{d_xd_t}\right)\bbeta\right]\right\}\cdot\\
&\times\exp\left\{-\frac{\Mqisigma}{2}\left[(\vecy- \matU\Mqetaz)^T\MqBxi\matT_0 - \Mqdelta^T\E_{-\bbeta}(\matZ_p^T\matZ_0)\right]\bbeta\right\},
\end{align*}
where 
\begin{align*}
\E(\matZ_j^T\matZ_k)&=\E\left[\sum_{i=1}^N(\matT_j^T\vecbxi)(\bxit\matT_k)\right]=\matT_j^T\left[\sum_{i=1}^N\E_{\bxi}(\vecbxi\bxit)\right]\matT_k,\qquad j,k=0,p.
\end{align*}
Thus, $q^*(\bbeta)=N(\Mqbeta,\Sqbeta)$ with
\begin{align*}
\Sqbeta&=\left\{\matT_0^T\left[\sum_{i=1}^N\E_{\bxi}\left(\vecbxi\bxit\right)\right]\matT_0\Mqisigma+\frac{1}{\sigma^2_{\bbeta}}\matI_{d_xd_t}\right\}^{-1}\\
\Mqbeta&=\Sqbeta\matT_0^T\left\{\MqBxit(\vecy- \matU\Mqetaz) - \left[\sum_{i=1}^N\E_{\bxi}\left(\vecbxi\bxit\right)\right]\matT_p\Mqdelta\right\}\Mqisigma.
\end{align*}
The derivation for $\bdelta$ is analogous and given by $q^*(\bdelta)=N(\Mqdelta,\Sqdelta)$ with
\begin{align*}
\Sqdelta&=\left\{\matT_p^T\left[\sum_{i=1}^N\E_{\bxi}\left(\vecbxi\bxit\right)\right]\matT_p\Mqisigma+\Mqlx\bPsi_x+\Mqlt\bPsi_t\right\}^{-1}\\
\Mqdelta&=\Sqdelta\matT_p^T\left\{\MqBxit(\vecy- \matU\Mqetaz) - \left[\sum_{i=1}^N\E_{\bxi}\left(\vecbxi\bxit\right)\right]\matT_0\Mqbeta\right\}\Mqisigma.
\end{align*}
\begin{align*}
\intertext{For $\sigma_x^2$, we have,} \sigma_x^2\mid\cdot &\sim
\text{IG}\left(\sum_{i=1}^Nn_i/2+a_x,\ b_x+\frac{1}{2}\sum_{i=1}^N\left[\tilde{\vecx}_i-\mu_x(\vect_i)-\Phi(\vect_i)\bxi_i\right]^T\left[\tilde{\vecx}_i-\mu_x(\vect_i)-\Phi(\vect_i)\bxi_i\right]\right)\\
\intertext{so that} q^*(\sigma^2_x)&\propto \exp\left\{-\Big(a_x+\sum_{i=1}^Nn_i/2-1\Big)\log(\sigma^2_x)-\frac{1}{\sigma_x^2}\left[b_x+
\frac{1}{2}\E_{-\sigma_x^2}\left(\sum_{i=1}^N\lVert\tilde{\vecx}_i-\mu_x(\vect_i)-\Phi(\vect_i)\bxi_i\rVert^2_2
\right)\right]\right\}.
\end{align*}
Therefore, $q^*(\sigma_x^2)=\text{IG}(a_x+\sum_{i=1}^Nn_i/2,\ \Bqsigmax),$ where
\[
\Bqsigmax=b_x+\frac{1}{2}\sum_{i=1}^N\left[\lVert\tilde{\vecx}_i-\mu_x(\vect_i)-\Phi(\vect_i)\Mqxi{i}\rVert^2_2+\trace\left(\Phi(\vect_i)^T\Phi(\vect_i)\Sqxi\right)\right]
\]
Note that for $\theta=\text{IG}(A,B),\ \mu_\theta(1/\theta)=A/B$.

Similarly,
\begin{align*}
p(\sigma^2\mid\cdot) &\propto (\sigma^2)^{-N/2 - a_s -1}\exp\left(-\frac{b_s + \frac{1}{2} \left(\vecy - \matU\boldeta_0-\boldeta_1\right)^T \left(\vecy - \matU\boldeta_0-\boldeta_1\right)}{\sigma^2} \right) \\
\text{so that } \sigma^2\mid\cdot &\sim \text{IG}\left(a = N/2 +
a_s,\ b = b_s + \frac{1}{2} \left(\vecy - \matU\boldeta_0-\boldeta_1\right)^T\left(\vecy - \matU\boldeta_0-\boldeta_1\right)\right)\\
\intertext{Thus,} q^*(\sigma^2)&\propto \exp\left\{-(a_s+N/2-1)\log(\sigma^2)-\frac{1}{\sigma^2}\left[b_s+
\frac{1}{2}\E_{-\sigma^2}\left(\lVert\left(\vecy - \matU\boldeta_0-\boldeta_1\right)\rVert^2_2
\right)\right]\right\}.
\end{align*}
\begin{align*}
\E_{-\sigma^2}\left[\lVert\left(\vecy - \matU\boldeta_0-\boldeta_1\right)\rVert^2_2\right]&=\E_{-\sigma^2}\left[\lVert(\vecy - \matU\Mqetaz-\Mqetao)\rVert^2_2\right]+
\E_{-\sigma^2}\left[\lVert\matU\boldeta_0-\matU\Mqetaz\rVert^2_2\right]\\
&+\E_{-\sigma^2}\left[\lVert\boldeta_1-\Mqetao\rVert^2_2\right].
\end{align*}
Now $\E_{-\sigma^2}\left[\lVert\matU\boldeta_0-\matU\Mqetaz\rVert^2_2\right]=\trace\left(\matU^T\matU\Sqetaz\right)$ and for the third term on the RHS we have
\begin{align*}
\E_{-\sigma^2}\lVert\boldeta_1&-\Mqetao\rVert^2_2]=\E_{-\sigma^2}\left[\sum_{i=1}^N(\bxit\btheta-\Mqbxi^T\Mqtheta)^2\right]=
\E_{-\sigma^2}\left[\sum_{i=1}^N\btheta^T\vecbxi\bxit\btheta\right]\\
&-\Mqtheta^T\Mqbxi^T\Mqbxi\Mqtheta=\E_{-\bxi_i}\left[\trace\left(\sum_{i=1}^N\vecbxi\bxit\Sqtheta\right)+\Mqtheta^T\sum_{i=1}^N\vecbxi\bxit\Mqtheta\right]\\
&-\Mqtheta^T\Mqbxi^T\Mqbxi\Mqtheta=\trace\left[\sum_{i=1}^N\E_{\bxi}\left(\vecbxi\bxit\right)\Sqtheta\right]+
\Mqtheta^T\left[\sum_{i=1}^N\E_{\bxi}\left(\vecbxi\bxit\right)\right]\Mqtheta\\
&-\Mqtheta^T\Mqbxi^T\Mqbxi\Mqtheta,
\end{align*}
where, as before, $\btheta=\matT(\bbeta^T,\bdelta^T)^T$.

Therefore, we have, $q^*(\sigma^2)=\text{IG}(a_s+N/2,\ \Bqsigma),$ where
\begin{align*}
\Bqsigma&=b_s+\frac{1}{2}\lVert(\vecy - \matU\Mqetaz-\Mqetao)\rVert^2_2
+\frac{1}{2}\trace\left(\matU^T\matU\Sqetaz\right)
+\frac{1}{2}\trace\left\{\left[\sum_{i=1}^N\E_{\bxi}\left(\vecbxi\bxit\right)\right]\bSigma_{q(\btheta)}\right\}\\
&+\frac{1}{2}\Mqtheta^T\left[\sum_{i=1}^N\E_{\bxi}\left(\vecbxi\bxit\right)\right]\Mqtheta-\frac{1}{2}\Mqtheta^T\Mqbxi^T\Mqbxi\Mqtheta.
\end{align*}


\subsection*{Laplace Approximation for Optimal Density for Principal Components}
First, defining some notation, the derivatives of the matrix valued
function $\vecM:\mathbb{R}^p\rightarrow\mathbb{R}^{m\times n}$ with
respect to $v_i$, $\vecv^T=(v_1,\ldots,v_p)$ and $\vecv$ are
\[
\mathcal{D}_{v_i}\vecM(\vecv)\equiv \left[\begin{array}{ccc} \frac{\partial m_{11}}{v_i} & \cdots & \frac{\partial m_{1n}}{v_i}\\
\vdots & \ddots & \vdots\\
\frac{\partial m_{m1}}{v_i} & \cdots & \frac{\partial m_{mn}}{v_i}
\end{array}\right],\qquad
\mathcal{D}_{\vecv^T}\vecM\equiv\left[\mathcal{D}_{v_1}\vecM\mid\cdots\mid\mcD_{v_p}\vecM\right],\qquad
\mathcal{D}_{\vecv}\vecM\equiv\left[\begin{array}{c}\mathcal{D}_{v_1}\vecM\\ \vdots\\ \mathcal{D}_{v_p}\vecM\end{array}\right]
\]
respectively.  Also define $\mcD_{\vecv^2}\vecM\equiv\mcD_{\vecv}(\mcD_{\vecv}\vecM)$.  We first differentiate the components of $\log q(\bxi_i)$ with respect to $\bxi_i$
\begin{align*}
\mcD_{\vecbxi}\trace&(\vecbxi\bxit\Sqtheta)
=\mcD_{\vecbxi}\trace(\vecbxi\bxit\Sqtheta)
=\mcD_{\vecbxi}\trace(\bxit\Sqtheta\vecbxi)
=\mcD_{\vecbxi}\bxit\Sqtheta\vecbxi=2\Sqtheta\vecbxi
\end{align*}
We then have, see, e.g., \textcite{vetter1973matrix},
\[
\mcD_{\bxi}\trace(\vecbxi\bxit\Sqtheta)=\mcD_{\bxi_i}(\bxit)\mcD_{\vecbxi}\trace(\vecbxi\bxit\Sqtheta)=2\mcD_{\bxi_i}(\bxit)\Sqtheta\vecbxi.
\]
\[
\mcD_{\bxi_i}(y_i- \vecu_i^T\Mqetaz-\bxit\Mqtheta)^2=-2\mcD_{\bxi_i}(\bxit)\Mqtheta(y_i-\vecu_i^T\Mqetaz-\bxit\Mqtheta)
\]
\[
\mcD_{\bxi_i}\left\{\bxi_i^T[\Mqisigmax\bPhi(\vect_i)^T\bPhi(\vect_i)+\diag(\bnu^{-1})]\bxi_i\right\}=2[\Mqisigmax\bPhi(\vect_i)^T\bPhi(\vect_i)+\diag(\bnu^{-1})]\bxi_i
\]
We arrive at
\begin{align*}
\mcD_{\bxi_i}\log q(\bxi_i)&=\Mqisigma\mcD_{\bxi_i}(\bxit)\Mqtheta(y_i-\vecu_i^T\Mqetaz-\bxit\Mqtheta)+\Mqisigmax(\tilde{\vecx}_i-\bmu_x(\vect_i))^T\bPhi(\vect_i)\\
&\qquad-[\Mqisigmax\bPhi(\vect_i)^T\bPhi(\vect_i)+\diag(\bnu^{-1})]\bxi_i-\Mqisigma\mcD_{\bxi_i}(\bxit)\Sqtheta\vecbxi
\end{align*}
Now to compute $\mcD_{\bxi_i^2}\log q(\bxi_i)$:
\begin{align*}
\mcD_{\bxi_i}&\left[\mcD_{\bxi_i}(\bxit)\Mqtheta(y_i-\vecu_i^T\Mqetaz-\bxit\Mqtheta)\right]=\mcD_{\bxi_i^2}(\bxit)\Mqtheta(y_i-\vecu_i^T\Mqetaz-\bxit\Mqtheta)\\
&-[\matI_M\otimes\mcD_{\bxi_i}(\bxit)\Mqtheta]\mcD_{\bxi_i}(\bxit)\Mqtheta=\mcD_{\bxi_i^2}(\bxit)\Mqtheta(y_i-\vecu_i^T\Mqetaz-\bxit\Mqtheta)\\
&-\Vec\left\{\mcD_{\bxi_i}(\bxit)\Mqtheta[\mcD_{\bxi_i}(\bxit)\Mqtheta]^T\right\}
\end{align*}
\[
\mcD_{\bxi_i}\left\{ [\Mqisigmax\bPhi(\vect_i)^T\bPhi(\vect_i)+\diag(\bnu^{-1})]\bxi_i\right\}=\Vec[\Mqisigmax\bPhi(\vect_i)^T\bPhi(\vect_i)+\diag(\bnu^{-1})]
\]
\begin{align*}
\mcD_{\bxi_i}\left[\mcD_{\bxi_i}(\bxit)\Sqtheta\vecbxi\right]&=\mcD_{\bxi_i^2}(\bxit)\Sqtheta\vecbxi
+[\matI_M\otimes\mcD_{\bxi_i}(\bxit)](\matI_M\otimes\Sqtheta)\mcD_{\bxi_i}(\vecbxi)\\
&=\mcD_{\bxi_i^2}(\bxit)\Sqtheta\vecbxi+[\matI_M\otimes\mcD_{\bxi_i}(\bxit)\Sqtheta]\mcD_{\bxi_i}(\vecbxi)\\
&=\mcD_{\bxi_i^2}(\bxit)\Sqtheta\vecbxi+[\matI_M\otimes\mcD_{\bxi_i}(\bxit)\Sqtheta]\Vec\left[\mcD^T_{\bxi_i}(\bxit)\right]\\
&=\mcD_{\bxi_i^2}(\bxit)\Sqtheta\vecbxi+\Vec\left[\mcD_{\bxi_i}(\bxit)\Sqtheta\mcD^T_{\bxi_i}(\bxit)\right],
\end{align*}
where $\otimes$ denotes the Kronecker product and the last equality
follows from, e.g., \textcite[Eq.~(9)]{vetter1973matrix}.  Thus, we
have
\begin{align}
\mcD_{\bxi_i^2}\log q&(\bxi_i)=\Mqisigma\mcD_{\bxi_i^2}(\bxit)\Mqtheta(y_i-\vecu_i^T\Mqetaz-\bxit\Mqtheta)\nonumber\\
&-\Mqisigma\Vec\left\{\mcD_{\bxi_i}(\bxit)\Mqtheta[\mcD_{\bxi_i}(\bxit)\Mqtheta]^T\right\}
-\Vec[\Mqisigmax\bPhi(\vect_i)^T\bPhi(\vect_i)+\diag(\bnu^{-1})]\nonumber\\
&-\Mqisigma\left\{\mcD_{\bxi_i^2}(\bxit)\Sqtheta\vecbxi+\Vec\left[\mcD_{\bxi_i}(\bxit)\Sqtheta\mcD^T_{\bxi_i}(\bxit)\right]\right\}
\end{align}
Next to derive expressions for $\mcD_{\bxi_i}(\bxit)$ and
$\mcD_{\bxi_i^2}(\bxit)$.  Let $\vecc(\bxi_i)=\bmu_x+\bPhi\bxi_i$ and
let $\Bxid$ be the $T\times K_xK_t$ matrix of derivatives of the
tensor product B-splines evaluated at $\vecc(\bxi_i)$ with $j$th row
denoted by $(\vecB')^T_{j,i}$.  Similarly, define $\Bxidd$, then
\[
\mcD_{\bxi_i}(\bxit)=\mcD_{\bxi_i}(\vecc^T)\mcD_\vecc\bxit=\mcD_{\bxi_i}(\vecc^T)\mcD_{\bxi_i}(\vecL^T\vecB_{\xi_i})=\bPhi^T\Bxid\odot(\vecL\otimes\vecone^T_{K_xK_t})
\]
and
\begin{align*}
\mcD_{\bxi_i^2}(\bxit)&=[\matI_M\otimes\bPhi^T]\mcD_{\bxi_i}[\Bxid\odot(\vecL\otimes\vecone^T_{K_xK_t})]=
(\matI_M\otimes\bPhi^T)(\mcD_{\bxi_i}(\vecc^T)\otimes\matI_T)\mcD_{\vecc}[\Bxid\odot(\vecL\otimes\vecone^T_{K_xK_t})]\\
&=(\matI_M\otimes\bPhi^T)(\bPhi^T\otimes\matI_T)\left(
                \begin{array}{c}
                  \ell_1\cdot(\vecB'')^T_{1,i} \\
                  \veczero_{T\times K_xK_t} \\
                  \ell_2\cdot(\vecB'')^T_{2,i} \\
                  \vdots \\
                  \ell_T\cdot(\vecB'')^T_{T,i} \\
                \end{array}
              \right)
=(\bPhi^T\otimes\bPhi^T)\left(
                \begin{array}{c}
                  \ell_1\cdot(\vecB'')^T_{1,i} \\
                  \veczero_{T\times K_xK_t} \\
                  \ell_2\cdot(\vecB'')^T_{2,i} \\
                  \vdots \\
                  \ell_T\cdot(\vecB'')^T_{T,i} \\
                \end{array}
              \right)
,
\end{align*}
where $\veczero_{m\times n}$ denotes a $m\times n$ matrix with every entry equal to 0.
Thus, we arrive at our Laplace approximation \ref{laplace}.

Next, we compute the expectations with respect to $\bxi_i$ involving
$\vecbxi$.  We use a second order matrix Taylor expansion about
$\bxi_{i,0}$.  Let $\tbxi=\bxi_i-\bxi_{i,0}$, we have
\begin{align*}
\vecbxi&\approx \bxiz+\mcD_{\bxi_i}[\bxiz]\tbxi+\frac{1}{2}\mcD_{\bxi_i^{T^2}}[\bxiz](\tbxi\otimes\tbxi)
\end{align*}
where
$\mathcal{D}^2_{\xi_i^{T^2}}[\bxiz]\equiv\mcD_{\xi_i^T}\{\mcD_{\xi_i^T}[\bxiz]\}$
with dimension $K_xK_t\times M^2$, see \textcite{vetter1973matrix}.
Therefore, we have
\[
\Mqbxi\approx\bxiz+\frac{1}{2}\mcD_{\bxi_i^{T^2}}[\bxiz]\Vec(\bLambda)=\bxiz+\frac{1}{2}\left\{\mcD_{\bxi_i^2}[\bxitz]\right\}^T\Vec(\bLambda)
\]
and
\begin{align*}
\vecbxi\bxit&\approx\left\{\bxiz+\mcD_{\bxi_i}[\bxiz]\tbxi+\frac{1}{2}\mcD_{\bxi_i^{T^2}}[\bxiz](\tbxi\otimes\tbxi) \right\}\\
&\qquad\qquad\qquad\times\left\{\bxiz+\mcD_{\bxi_i}[\bxiz]\tbxi+\frac{1}{2}\mcD_{\bxi_i^{T^2}}[\bxiz](\tbxi\otimes\tbxi) \right\}^T\\
&=\bxiz\bxitz+\bxiz\tbxi^T\mcD_{\bxi_i}^T[\bxiz]+\mcD_{\bxi_i}[\bxiz]\tbxi\bxitz\\
&+\frac{1}{2}\bxiz(\tbxi^T\otimes\tbxi^T)\mcD_{\bxi_i^2}[\bxitz]+\frac{1}{2}\mcD_{\bxi_i^{T^2}}[\bxiz](\tbxi\otimes\tbxi)\bxitz\\
&+\mcD_{\bxi_i}[\bxiz]\tbxi\tbxi^T\mcD_{\bxi^T_i}[\bxitz]+o(\lVert\tbxi\rVert^2)
\end{align*}
so that
\begin{align*}
\E_{\bxi_i}[\vecbxi\bxit]&\approx\bxiz\bxitz+\frac{1}{2}\bxiz\Vec(\bLambda)^T\mcD_{\bxi_i^2}(\bxitz)+\frac{1}{2}\mcD_{\bxi_i^{T^2}}(\bxiz)\Vec(\bLambda)\bxitz\\
&+\mcD_{\bxi_i^T}[\bxiz]\bLambda\mcD_{\bxi_i}[\bxitz]=\bxiz\bxitz+\bxiz\Vec(\bLambda)^T\mcD_{\bxi_i^2}(\bxitz)\\
&+\left\{\mcD_{\bxi_i}[\bxitz]\right\}^T\bLambda\mcD_{\bxi_i}[\bxitz]
\end{align*}
\section{Derivation Of Log-Likelihood Lower Bound}\label{app_llbound}
For any density,
$q^*$, a lower bound on our log-likelihood can be derived using
Kullbeck-Leibler divergence and is given by
\autocite[e.g.,][]{ormerod2010explaining}
\[
\log[p(\vecy,\tilde{\vecx};\bTheta)]\geq\log[\underline{p}(\vecy,\tilde{\vecx};q)]:=\!\int q^*(\bTheta)\log\left(\frac{p(\vecy,\tilde{\vecx},\bTheta)}{q^*(\bTheta)}\right)\!d\bTheta=
\E_{q^*}\{\log[p(\vecy,\tilde{\vecx},\bTheta)]-\log[q^*(\bTheta)]\}.
\]
For FGAM \eqref{fullmodel} we have
\begin{align}\label{logliklb}
\log &[\underline{p}(\vecy,\tilde{\vecx};q)]=\Eq\{\log[p(\vecy\mid\boldeta_0,\bbeta,\bdelta,\bXi,\sigma^2)]\}+\Eq\{\log[p(\tilde{\vecx}\mid\bXi,\sigma_x^2)]\}\nonumber\\
&+\Eq\{\log[p(\boldeta_0)]-\log[q^*(\boldeta_0)]\}+\Eq\{\log[p(\bbeta)]-\log[q^*(\bbeta)]\}+\Eq\{\log[p(\bdelta)]-\log[q^*(\bdelta)]\}\nonumber\\
&+\text{$\sum_{i=1}^N$}\Eq\{\log[p(\bxi_i)]-\log[q^*(\bxi_i)]\}+\Eq\{\log[p(\lambda_x)]-\log[q^*(\lambda_x)]\}\nonumber\\
&+\Eq\{\log[p(\lambda_t)]-\log[q^*(\lambda_t)]\}+\Eq\{\log[p(\sigma^2)]-\log[q^*(\sigma^2)]\}+\Eq\{\log[p(\sigma_x^2)]-\log[q^*(\sigma_x^2)]\}
\end{align}

The first term in (\ref{logliklb}) is
\begin{align*}
\Eq\{\log [p(\vecy\mid\boldeta_0,\bbeta,\bdelta,\bXi,\sigma^2)]\}&=\Eq\left[-\frac{N}{2}\log(\sigma^2)-\frac{1}{2\sigma^2}\lVert\vecy - \matU\boldeta_0-\boldeta_1\rVert^2_2\right]+C\\
&=-\frac{N}{2}\Eq[\log(\sigma^2)]-\Mqisigma(\Bqsigma-b_s)+C,
\end{align*}
where $C$ is used from here on to represent any constant that will not
affect the log-likelihood as the parameter estimates are updated.  The
second term in (\ref{logliklb}) is
\begin{align*}
\Eq\{\log[p(\tilde{\vecx}\mid\bXi,\sigma_x^2)]\}&=\Eq\left[-\frac{\sum_{i=1}^Nn_i}{2}\log(\sigma^2_x)-\frac{1}{2\sigma^2_x}\sum_{i=1}^N\lVert\tilde{\vecx}_i-\mu_x(\vect_i)-\Phi(\vect_i)\bxi_i\rVert^2_2\right]+C\\
&=-\frac{\sum_{i=1}^Nn_i}{2}\Eq[\log(\sigma^2_x)]-\Mqisigmax(\Bqsigmax-b_x)+C.
\end{align*}

The third term (recalling that $\sigma^2_{\boldeta_0}$ is fixed) is
\begin{align*}
\Eq\{&\log[p(\boldeta_0)]-\log[q^*(\boldeta_0)]\}=\Eq\left[-\frac{1}{2\sigma^2_{\boldeta_0}}\boldeta_0^T\boldeta_0+\frac{1}{2}\log(\lvert\Sqetaz\rvert)\right.\\
&+\left.\frac{1}{2}(\boldeta_0-\Mqetaz)^T\Sqetaz^{-1}(\boldeta_0-\Mqetaz) \right]+C\\
&=-\frac{1}{2\sigma^2_{\boldeta_0}}\left[\Mqetaz^T\Mqetaz+\trace(\Sqetaz)\right]+\frac{1}{2}\log(\lvert\Sqetaz\rvert)+C
\end{align*}

The fourth term (recalling that $\sigma^2_{\bbeta}$ is fixed) is
\begin{align*}
\Eq\{\log[p(\bbeta)]-\log[q^*(\bbeta)]\}&=\Eq\left[-\frac{1}{2\sigma^2_{\bbeta}}\bbeta^T\bbeta+\frac{1}{2}\log(\lvert\Sqbeta\rvert)+\frac{1}{2}(\bbeta-\Mqbeta)^T\Sqbeta^{-1}(\bbeta-\Mqbeta) \right]\\
&+C=-\frac{1}{2\sigma^2_{\bbeta}}\left[\Mqbeta^T\Mqbeta+\trace(\Sqbeta)\right]+\frac{1}{2}\log(\lvert\Sqbeta\rvert)+C
\end{align*}

The fifth term is
\begin{align*}
\Eq\{\log&[p(\bdelta)]-\log[q^*(\bdelta)]\}=\Eq\left[\frac{1}{2}\log\lvert\lambda_x \bPsi_x +  \lambda_t \bPsi_t\rvert-\frac{1}{2}\bdelta^T (\lambda_x \bPsi_x+\lambda_t \bPsi_t)\bdelta\right.\\
&\qquad+\left.\frac{1}{2}\log(\lvert\Sqdelta\rvert)+\frac{1}{2}(\bdelta-\Mqdelta)^T\Sqdelta^{-1}(\bdelta-\Mqdelta) \right]+C\\
&\leq\frac{1}{2}\log\lvert\Mqlx \bPsi_x +  \Mqlt \bPsi_t\rvert-\frac{1}{2}\Mqdelta^T(\Mqlx \bPsi_x +  \Mqlt \bPsi_t)\Mqdelta\\
&\qquad-\frac{1}{2}\Mqlx\trace\left(\bPsi_x\Sqdelta\right)-\frac{1}{2}\Mqlt\trace\left(\bPsi_t\Sqdelta\right)+\frac{1}{2}\log(\lvert\Sqdelta\rvert)+C
\end{align*}
Where the inequality follows from Jensen's inequality and the
log-concavity of the determinant over the class of positive definite
matrices.  This inequality is not in the direction we want.  If we use
the approximation
$\Eq\log\lvert\lambda_x \bPsi_x + \lambda_t
\bPsi_t\rvert\approx\log\lvert\Mqlx \bPsi_x + \Mqlt \bPsi_t\rvert$, we
appear to lose our guarantee of increasing the lower bound on the
log-likelihood at each iteration.

In the sixth term we have
\begin{align*}
\Eq\{\log[&p(\bxi_i)]-\log[q^*(\bxi_i)]\}=\Eq\left[-\frac{1}{2}\bxi_i^T\diag(\bnu^{-1})\bxi_i+\frac{M}{2}\log(\lvert\bLambda_i\rvert)+\frac{1}{2}(\bxi_i-\bxi_{i,0})^T\bLambda_i^{-1}(\bxi_i-\bxi_{i,0}) \right]\\
&+C=-\frac{1}{2}\left\{\bxi_{i,0}^T\diag(\bnu^{-1})\bxi_{i,0}+\trace[\diag(\bnu^{-1})\bLambda_i]\right\}+\frac{M}{2}\log(\lvert\bLambda_i\rvert)+C;\ i=1,\ldots,N.
\end{align*}

For the seventh term
\begin{align*}
\Eq\{\log[p(\lambda_x)]-&\log[q^*(\lambda_x)]\}=\Eq\left\{(a_l+1)\log(\lambda_x)-b_l\lambda_x-\frac{1}{2}\log\lvert\lambda_x \bPsi_x +  \Mqlt \bPsi_t\rvert\right.\\
&\left. -\log(c_{q(\lambda_x)})+\frac{1}{2} \left(\text{tr}(\bPsi_x\Sqdelta)+ \Mqdelta^T\bPsi_x\Mqdelta\right)\lambda_x-(a_l + 1)\log(\lambda_x) +b_l \lambda_x\right\}+C\\
&\approx(a_l+1)\Eq[\log(\lambda_x)]-\frac{1}{2}\log\lvert\Mqlx \bPsi_x +  \Mqlt \bPsi_t\rvert-\log(c_{q(\lambda_x)})\\
&\qquad+\frac{1}{2} \left(\text{tr}(\bPsi_x\Sqdelta)+ \Mqdelta^T\bPsi_x\Mqdelta\right)\Mqlx+C
\end{align*}

For the eighth term
\begin{align*}
\Eq\{\log[p(\lambda_t)]-\log[q^*(\lambda_t)]\}&\approx(a_l+1)\Eq[\log(\lambda_t)]-\frac{1}{2}\log\lvert\Mqlx \bPsi_x +  \Mqlt \bPsi_t\rvert-\log(c_{q(\lambda_t)})\\
&\qquad+\frac{1}{2} \left(\text{tr}(\bPsi_t\Sqdelta)+ \Mqdelta^T\bPsi_t\Mqdelta\right)\Mqlt+C
\end{align*}

For the ninth term
\begin{align*}
\Eq\{\log[p(\sigma^2)]-\log[q^*(\sigma^2)]\}&=\Eq\left\{-(a_s+1)\log(\sigma^2)-\frac{b_s}{\sigma^2}-(a_s+N/2)\log(\Bqsigma)\right.\\
&\qquad+\left. (a_s+N/2+1)\log(\sigma^2)+\frac{\Bqsigma}{\sigma^2}\right\}+C\\
&=\frac{N}{2}\Eq[\log(\sigma^2)]-(a_s+N/2)\log(\Bqsigma)+\Mqisigma(\Bqsigma-b_s)+C
\end{align*}
The tenth term is
\begin{align*}
\Eq\{\log[p(\sigma_x^2)]-\log&[q^*(\sigma_x^2)]\}=\Eq\left\{-(a_x+1)\log(\sigma_x^2)-\frac{b_x}{\sigma_x^2}-(a_x+\sum_{i=1}^Nn_i/2)\log(\Bqsigmax)\right.\\
&\qquad\qquad+\left. (a_x+\sum_{i=1}^Nn_i/2+1)\log(\sigma_x^2)+\frac{\Bqsigmax}{\sigma_x^2}\right\}+C\\
&=\frac{\sum_{i=1}^Nn_i}{2}\Eq[\log(\sigma_x^2)]-(a_x+\sum_{i=1}^Nn_i/2)\log(\Bqsigmax)+\Mqisigmax(\Bqsigmax-b_x)+C
\end{align*}

Combining all ten terms, several components cancel and we are left with
\begin{align}\label{loglik}
\log [\underline{p}(\vecy,\tilde{\vecx}&;q)]\approx
-\frac{1}{2\sigma^2_{\boldeta_0}}\left[\Mqetaz^T\Mqetaz+\trace(\Sqetaz)\right]+\frac{1}{2}\log(\lvert\Sqetaz\rvert)
-\frac{1}{2\sigma^2_{\bbeta}}\left[\Mqbeta^T\Mqbeta+\trace(\Sqbeta)\right]\notag\\
&+\frac{1}{2}\log(\lvert\Sqbeta\rvert)
-\frac{1}{2}\sum_{i=1}^N\left\{\bxi_{i,0}^T\diag(\bnu^{-1})\bxi_{i,0}+\trace[\diag(\bnu^{-1})\bLambda_i]-M\log(\lvert\bLambda_i\rvert)\right\}\notag\\
&+(a_l+1)\Eq[\log(\lambda_x)]-(a_s+N/2)\log(\Bqsigma)-(a_x+\sum_{i=1}^Nn_i/2)\log(\Bqsigmax)\notag\\
&+\frac{1}{2}\log(\lvert\Sqdelta\rvert)+(a_l+1)\Eq[\log(\lambda_t)]-\frac{1}{2}\log\lvert\Mqlx \bPsi_x +  \Mqlt \bPsi_t\rvert-\log(c_{q(\lambda_t)}/c_{q(\lambda_x)})
\end{align}
\section{Complete Variational Bayes Algorithm}\label{app_vbalg}
Below is the full VB algorithm.  Note that it is spread over two pages.
\begin{algorithm}
\caption{Steps for estimating parameters from optimal densities, $q^*(\btheta)$, for FGAM}
\label{VBalgorithm}
\begin{algorithmic} [1]
\State Initialize $\Bqsigma,\Bqsigmax,\Mqlx,\Mqlt>0$, $\Sqetaz=\matI_{p_0},\ \Sqbeta=\matI_{d_xd_t},\ \Sqdelta=\matI_{K_xK_t-d_xd_t},$
\Statex $\qquad\qquad\ \Mqetaz=\veczero,\ \Mqbeta=\veczero,\ \Mqdelta=\veczero$.
\State Choose grid of $G$ points, $\vecg$, and obtain Gauss-Laguerre quadrature weights, $\vecL_g$, for numerical integration of optimal densities for $\lambda_x,\ \lambda_t$.
\State Compute $\bnu,\ \bmu_x,\ \bPhi$, $\bmu_x(\vect_i)$, $\bPhi(\vect_i),\ i=1,\ldots,N$, from an initial functional principal components analysis.
\Repeat
\For{$i = 1 \to N$}
\State $\bxi_{i,0}\gets\text{mode of }\log q(\bxi_i)=\Mqisigma\!\left[(y_i-\vecu_i^T\Mqetaz)\bxit\Mqtheta-\frac{1}{2}(\bxit\Mqtheta)^2+\frac{1}{2}\bxit\Sqtheta\vecbxi\right]$
\Statex $\qquad\qquad\qquad\qquad\qquad\qquad+(\tilde{\vecx}_i-\bmu_x(\vect_i))^T\bPhi(\vect_i)\bxi_i-\frac{1}{2}\bxi_i^T\left[\Mqisigmax\bPhi^T(\vect_i)\bPhi(\vect_i)+\diag(\bnu^{-1})\right]\bxi_i$
\State $\mcD_{\bxi_i}[\bxitz]\gets\bPhi^T\vecB'_{\bxi_{i,0}}\odot(\vecL\otimes\vecone^T_{K_xK_t})$
\State $\mcD_{\bxi^2_i}[\bxitz]\gets(\bPhi^T\otimes\bPhi^T)\left[\ell_1\cdot(\vecB''_{\bxi_{i,0}})_{1,i},\veczero_{K_xK_t\times T},\ell_2\cdot(\vecB''_{\bxi_{i,0}})_{2,i},\ldots,\ell_T\cdot(\vecB''_{\bxi_{i,0}})_{T,i}\right]^T$
\State $\Vec(\bLambda_i^{-1})\gets\Big[\Mqisigma\mcD_{\bxi_i^2}(\bxit)\Mqtheta(y_i-\vecu_i^T\Mqetaz-\bxit\Mqtheta)$
\Statex $\ \ \qquad\qquad+\Mqisigma\Vec\left\{\mcD_{\bxi_i}(\bxit)\Mqtheta[\mcD_{\bxi_i}(\bxit)\Mqtheta]^T\right\}+\Vec[\Mqisigmax\bPhi(\vect_i)^T\bPhi(\vect_i)+\diag(\bnu^{-1})]$
\Statex $\ \ \qquad\qquad+\Mqisigma\left\{\mcD_{\bxi_i^2}(\bxit)\Sqtheta\vecbxi+\Vec\left[\mcD_{\bxi_i}(\bxit)\Sqtheta\mcD^T_{\bxi_i}(\bxit)\right]\right\}\Big]_{\bxi_i=\bxi_{i,0}}$
\State $\Mqbxi\gets\bxiz+\frac{1}{2}\left\{\mcD_{\bxi_i^2}[\bxitz]\right\}^T\Vec(\bLambda_i)$
\State $\E_{\bxi_i}[\vecbxi\bxit]\gets\bxiz\bxitz+\bxiz\Vec(\bLambda_i)^T\mcD_{\bxi_i^2}(\bxitz)$
\Statex $\qquad\qquad\qquad+\left\{\mcD_{\bxi_i}[\bxitz]\right\}^T\bLambda_i\mcD_{\bxi_i}[\bxitz]$
\EndFor
\State $\Sqetaz \gets \left\{\Mqisigma\matU^T\matU+\frac{1}{\sigma^2_{\eta_0}}\matI_{p_0}\right\}^{-1}$
\State $\Mqetaz \gets \Sqetaz\matU^T\left(\vecy-\Mqetao\right)\Mqisigma$

\algstore{myalg}
\end{algorithmic}
\end{algorithm}
\begin{algorithm}
\begin{algorithmic} [1]                   
\algrestore{myalg}
\State $\Sqbeta \gets \left\{\matT_0^T\left[\sum_{i=1}^N\E_{\bxi}\left(\vecbxi\bxit\right)\right]\matT_0\Mqisigma+\frac{1}{\sigma^2_{\bbeta}}\matI_{d_xd_t}\right\}^{-1}$
\State $\Mqbeta \gets \Sqbeta\matT_0^T\left\{\MqBxit(\vecy- \matU\Mqetaz) - \left[\sum_{i=1}^N\E_{\bxi}\left(\vecbxi\bxit\right)\right]\matT_p\Mqdelta\right\}\Mqisigma$
\State $\Sqdelta \gets \left\{\matT_p^T\left[\sum_{i=1}^N\E_{\bxi}\left(\vecbxi\bxit\right)\right]\matT_p\Mqisigma+\Mqlx\bPsi_x+\Mqlt\bPsi_t\right\}^{-1}$
\State $\Mqdelta \gets \Sqdelta\matT_p^T\left\{\MqBxit(\vecy- \matU\Mqetaz) - \left[\sum_{i=1}^N\E_{\bxi}\left(\vecbxi\bxit\right)\right]\matT_0\Mqbeta\right\}\Mqisigma$
\For{$i = 1 \to G$}
\State $\ell_{\lambda_x}(g_i)\gets\frac{1}{2}\log\lvert g_i \bPsi_x +  \Mqlt \bPsi_t\rvert-g_i\left\{b_l +\frac{1}{2} \left[\text{tr}(\bPsi_x\Sqdelta)+
\Mqdelta^T\bPsi_x\Mqdelta\right]\right\}$
\EndFor
\State $\Mqlx\gets[\vecL_g^T\ell_{\lambda_x}(\vecg)]^{-1}\vecL_g^T\left\{\vecg\odot\exp[\ell_{\lambda_x}(\vecg)-\max_\vecg\ell_{\lambda_x}(\vecg)]\right\}$
\For{$i = 1 \to G$}
\State $\ell_{\lambda_t}(g_i)\gets\frac{1}{2}\log\lvert\Mqlx \bPsi_x + g_i \bPsi_t\rvert-g_i\left\{b_l +\frac{1}{2} \left[\text{tr}(\bPsi_t\Sqdelta)+
\Mqdelta^T\bPsi_t\Mqdelta\right]\right\}$
\EndFor
\State $\Mqlt\gets[\vecL_g^T\ell_{\lambda_t}(\vecg)]^{-1}\vecL_g^T\left\{\vecg\odot\exp[\ell_{\lambda_t}(\vecg)-\max_\vecg\ell_{\lambda_t}(\vecg)]\right\}$
\State $\Bqsigmax\gets b_x+\frac{1}{2}\sum_{i=1}^N\left[\lVert\tilde{\vecx}_i-\mu_x(\vect_i)-\Phi(\vect_i)\bxi_{i,0}\rVert^2_2
+\trace\left(\Phi(\vect_i)^T\Phi(\vect_i)\bLambda_i\right)\right]$
\State $\Mqisigmax\gets (a_x+\sum_{i=1}^Nn_i/2)/\Bqsigmax$
\State $\Bqsigma\gets b_s+\frac{1}{2}\lVert(\vecy - \matU\Mqetaz-\Mqetao)\rVert^2_2+\frac{1}{2}\trace\left(\matU^T\matU\Sqetaz\right)+
\frac{1}{2}\trace\left\{\left[\sum_{i=1}^N\E_{\bxi}\left(\vecbxi\bxit\right)\right]\bSigma_{q(\btheta)}\right\}$
\Statex $\qquad\qquad+\frac{1}{2}\Mqtheta^T\left[\sum_{i=1}^N\E_{\bxi}\left(\vecbxi\bxit\right)\right]\Mqtheta
-\frac{1}{2}\Mqtheta^T\Mqbxi^T\Mqbxi\Mqtheta$
\State $\Mqisigma\gets (a_x+N/2)/\Bqsigma$
\Until{Change in $\underline{p}(\vecy,\tilde{\vecx};q)$ is negligible \emph{OR} maximum number of iterations reached}
\end{algorithmic}
\end{algorithm}

%
\clearpage
\printbibliography
\end{document}